\documentclass[aps,prd,twocolumn,superscriptaddress,nofootinbib]{revtex4-1}

\usepackage{amsfonts}
\usepackage{psfrag}
\usepackage{graphicx}
\usepackage{cancel}
\usepackage{amsmath}
\usepackage{natbib}
\usepackage{xr}
\usepackage{hyperref}
\usepackage{cleveref}
\usepackage{verbatim}
\usepackage{multirow}

\usepackage{booktabs,dcolumn}

\newcolumntype{d}[1]{D{.}{.}{#1}}

\RequirePackage[normalem]{ulem} 
\RequirePackage{color}\definecolor{RED}{rgb}{1,0,0}\definecolor{BLUE}{rgb}{0,0,1} 
\providecommand{\DIFaddbegin}{} 
\providecommand{\DIFaddend}{} 
\providecommand{\DIFdelbegin}{} 
\providecommand{\DIFdelend}{} 
\providecommand{\DIFaddbeginFL}{} 
\providecommand{\DIFaddendFL}{} 
\providecommand{\DIFdelbeginFL}{} 
\providecommand{\DIFdelendFL}{} 
\newcommand{\DIFscaledelfig}{0.5}
\RequirePackage{settobox} 
\RequirePackage{letltxmacro} 
\newsavebox{\DIFdelgraphicsbox} 
\newlength{\DIFdelgraphicswidth} 
\newlength{\DIFdelgraphicsheight} 
\LetLtxMacro{\DIFOincludegraphics}{\includegraphics} 
\newcommand{\DIFaddincludegraphics}[2][]{{\color{blue}\fbox{\DIFOincludegraphics[#1]{#2}}}} 
\newcommand{\DIFdelincludegraphics}[2][]{
\sbox{\DIFdelgraphicsbox}{\DIFOincludegraphics[#1]{#2}}
\settoboxwidth{\DIFdelgraphicswidth}{\DIFdelgraphicsbox} 
\settoboxtotalheight{\DIFdelgraphicsheight}{\DIFdelgraphicsbox} 
\scalebox{\DIFscaledelfig}{
\parbox[b]{\DIFdelgraphicswidth}{\usebox{\DIFdelgraphicsbox}\\[-\baselineskip] \rule{\DIFdelgraphicswidth}{0em}}\llap{\resizebox{\DIFdelgraphicswidth}{\DIFdelgraphicsheight}{
\setlength{\unitlength}{\DIFdelgraphicswidth}
\begin{picture}(1,1)
\thicklines\linethickness{2pt} 
{\color[rgb]{1,0,0}\put(0,0){\framebox(1,1){}}}
{\color[rgb]{1,0,0}\put(0,0){\line( 1,1){1}}}
{\color[rgb]{1,0,0}\put(0,1){\line(1,-1){1}}}
\end{picture}
}\hspace*{3pt}}} 
} 
\LetLtxMacro{\DIFOaddbegin}{\DIFaddbegin} 
\LetLtxMacro{\DIFOaddend}{\DIFaddend} 
\LetLtxMacro{\DIFOdelbegin}{\DIFdelbegin} 
\LetLtxMacro{\DIFOdelend}{\DIFdelend} 
\DeclareRobustCommand{\DIFaddbegin}{\DIFOaddbegin \let\includegraphics\DIFaddincludegraphics} 
\DeclareRobustCommand{\DIFaddend}{\DIFOaddend \let\includegraphics\DIFOincludegraphics} 
\DeclareRobustCommand{\DIFdelbegin}{\DIFOdelbegin \let\includegraphics\DIFdelincludegraphics} 
\DeclareRobustCommand{\DIFdelend}{\DIFOaddend \let\includegraphics\DIFOincludegraphics} 
\LetLtxMacro{\DIFOaddbeginFL}{\DIFaddbeginFL} 
\LetLtxMacro{\DIFOaddendFL}{\DIFaddendFL} 
\LetLtxMacro{\DIFOdelbeginFL}{\DIFdelbeginFL} 
\LetLtxMacro{\DIFOdelendFL}{\DIFdelendFL} 
\DeclareRobustCommand{\DIFaddbeginFL}{\DIFOaddbeginFL \let\includegraphics\DIFaddincludegraphics} 
\DeclareRobustCommand{\DIFaddendFL}{\DIFOaddendFL \let\includegraphics\DIFOincludegraphics} 
\DeclareRobustCommand{\DIFdelbeginFL}{\DIFOdelbeginFL \let\includegraphics\DIFdelincludegraphics} 
\DeclareRobustCommand{\DIFdelendFL}{\DIFOaddendFL \let\includegraphics\DIFOincludegraphics} 

\newcommand{\e}{\mathrm{e}}
\newcommand{\D}{\mathrm{d}}
\newcommand{\U}{U_\mathrm{tot}}

\begin{document}

\title{Relaxation times for Bose-Einstein condensation in axion miniclusters}

\author{Kay Kirkpatrick}
\email{kkirkpat@illinois.edu}
\affiliation{Department of Mathematics,
University of Illinois at Urbana-Champaign
Urbana, IL 61801}
\author{Anthony E. Mirasola}
\email{aem8@illinois.edu}
\affiliation{Department of Physics,
University of Illinois at Urbana-Champaign
Urbana, IL 61801}
\author{Chanda Prescod-Weinstein}
\email{Chanda.Prescod-Weinstein@unh.edu}
\affiliation{Department of Physics \& Astronomy, University of New Hampshire, Durham, NH 03824}

\begin{abstract}
We study the Bose condensation of scalar dark matter in the presence of both gravitational and self-interactions. Axions and other scalar dark matter in gravitationally bound miniclusters or dark matter halos are expected to condense into Bose-Einstein condensates called Bose stars. This process has been shown to occur through attractive self-interactions of the axion-like particles or through the field's self gravitation. We show that in the high-occupancy regime of scalar dark matter, the Boltzmann collision integral does not describe either gravitaitonal or self-interactions, and derive kinetic equations valid for these interactions. We use this formalism to compute relaxation times for the Bose-Einstein condensation, and find that condensation into Bose stars could occur within the lifetime of the universe. The self-interactions reduce the condensation time only when they are very strong.
\end{abstract}

\maketitle

\section{Introduction}
The composition of dark matter is one of the most longstanding problems in cosmology. The dominant model, known as Lambda Cold Dark Matter ($\Lambda$CDM), proposes that the dark matter is cold and has a low velocity dispersion. It has been successful at cosmological distance scales \cite{PLANCK2018}. However, at galactic distance scales and smaller ($\lesssim 10$ kpc) it has a number of problems. At these scales, the predicted density profiles disagree with observations and a higher abundance of dwarf galaxies is predicted than is observed \cite{WeinbergControversies2015,CDMproblems,TooBigToFail}. While there are several proposed solutions to these problems \cite{AvilaReese2001,Kamionkowski2000,Spergel2000,Governato2010,BuckleyPeter2018}, an attractive proposal considers the quantum properties of the dark matter particles. In this case, the large-scale predictions remain the same as in $\Lambda$CDM, but on scales less than the de~Broglie wavelength the predictions change.

Among the proposed candidates for the dark matter are light bosons, such as the QCD axion \cite{PecceiQuinn,Weinberg1978,Wilczek1978,DineFischlerSrednicki,Turner1986,SikivieNotes} or fuzzy dark matter composed of ultra-light axions or other scalar fields \cite{StringAxiverse,WittenUltralight,Ringwald2012}. The QCD axion is especially well-motivated since it is hypothesized as a solution to the strong CP problem in QCD, and it has been shown that it could account for the correct dark matter abundance \cite{Preskill1983,Kim2010}. Both the QCD axion and ultralight scalars are the subjet of several ongoing experimental searches, most notably ADMX \cite{ADMX2020} and ABRACADABRA \cite{ABRACADABRA2019}. Experiments into the neutron electron dipole moment can also constrain models of axion physics \cite{NeutronDipole2020}.

These proposed candidates have in common that they thermalize to form compact, gravitationally bound solitons that can be described as Bose-Einstein condensation \cite{Lee1992,Jetzer1992,Kolb1993,SelfInteractionSign,SemikozTkachev,Khlebnikov2000,Sikivie2009,Erken2012}. This phenomenon is proposed to occur over a wide region of the parameter space of masses and self-couplings, with masses ranging from $m\sim 10^{-22}$ eV in the case of ultralight axions to $m\sim 10^{-5}$ eV in the case of QCD axions. 

While there have been many studies of the condensed objects formed by axions, these have focused on the equilibrium properties of these objects \cite{RuffiniBonazzola,Chavanis2011,Barranco2011,EbyGlobalView,EbySelfInteraction,EbyInfrared,EbyLifetime,EbyApprox,BraatenDense,Visinelli2018}. In the existing studies of the formation process, the long and short range interaction have rarely been considered together. This literature has shown that gravitational interaction can lead to relaxation times of axion-like particles into Bose condensed objects that are less than the lifetime of the observable universe \cite{SemikozTkachev,Khlebnikov2000,Sikivie2009,Schive2014,Levkov2018}.

This paper is organized as follows. In Sec.\ II we show that the axion dark matter is described by a classical non-relativistic scalar field, and show that statistical ensembles of these fields are described by a Wigner distribution. In Sec.\ III we expand the equation of motion to obtain timescales for the relaxation processes associated with the gravitational and attractive self-interactions. In Sec. IV we discuss the relation between these timescales and their implications on the relevance of gravitational and self-interactions in the thermalization of axion stars.

\section{Cosmological axions: A statistical ensemble of waves}
\subsection{Axion miniclusters}
There are two main scenarios that can generate structure formation of axion miniclusters or halos. For the QCD axion, the Peccei-Quinn symmetry breaking could occur after inflation. This happens when the symmetry breaking scale $f_a$ is less than the energy scale of inflation. For QCD axions this is possible since its symmetry breaking scale is lower, $f_a\sim 10^{12}$ GeV, but for ultralight axion-like particles, the symmetry breaking scale is too high, $f_a \sim 10^{16}$ GeV \cite{AxionCosmology}. 

In this scenario, the miniclusters are formed through the Kibble mechanism. Symmetry breaking causes the axion field to take random, uncorrelated values in different Hubble patches, resulting in density fluctuations of $\mathcal O(1)$ which decouple from the background Hubble expansion to form miniclusters \cite{HoganRees,Kolb1993,Nelson2018,Kibble1976}. Compared to the dark matter halos of $\Lambda$CDM, these are quite small with masses on the order of $10^{-13} M_\odot$, and radii on the order of $10^5$ km, determined by the mass and size of the horizon at the QCD phase transition \cite{MiniclusterSpectrum}. This distance scale also determines the characteristic wavenumber of the axions, since initially the de~Broglie wavelength of the axions is the size of the Hubble patch. However, the wavenumber is redshifted after the QCD phase transition. 

In the second scenario, CP symmetry can be broken during inflation. This scenario occurs when the symmetry breaking sale $f_a$ is higher than the energy scale inflation, which could be the case for QCD axions or ultralight scalars. In this scenario the axion field in our current universe originates from a single Hubble patch at the time of the symmetry breaking, and so does not exhibit $\mathcal O(1)$ fluctuations since these are inflated away. Fluctuations in the axion field that generate structure can still arise in a number of ways: by gravitational collapse due to the Jeans instability \cite{AxionCosmology}, by the growth of quantum fluctuations in the axion field enhanced by the axion's self-interactions \cite{Arvanitaki2020}, or by a phase transition in the sector determining the axion's mass \cite{Hardy2017}. Recently, Ref. \cite{Fukunaga2020} found that this scenario does not result in minicluster formation for the QCD axion, but can lead to minicluster formation for other axion like particles.

In this paper, we refer to any gravitationally bound structure of axion or ultralight scalar dark matter, formed pre- or post-inflation, as an axion minicluster. While these structures have different masses, sizes, and observational signatures, they all consist of scalar dark matter with the potential to form solitonic cores such as Bose stars, where the scalar field is in its ground state.

\subsection{Gross-Pitaevskii-Poisson Equations}

In this section, we review how the axions or scalar dark matter can be described by a classical complex field evolving under a system of equations known as the Gross-Pitaevskii-Poisson equations. Axion or scalar dark matter is a scalar field $\phi$. In the case of the QCD axion, it arises as the Goldstone boson for a spontaneously broken symmetry, which in the instanton approximation results in the following potential,
\begin{equation}
    V(\phi)=m^2 f_a ^2\left[1- \cos\frac{\phi}{f_a} \right].
    \label{axionpotential}
\end{equation}
Here $m$ is the mass of the axions and $f_a$ is the symmetry breaking scale of the axion. 
When $\phi\ll f_a$, as is the case for cosmological axions, we can expand the potential to fourth order to obtain
\begin{equation}
    V(\phi)=\frac{1}{2}m^2\phi^2 + \lambda \phi^4,
    \label{eq:fourthorder}
\end{equation}
where 
\begin{equation}
    \lambda =-\frac{1}{4!} \frac{m^2}{f_a ^2}
\end{equation} is the attractive quartic self-interaction. 

For QCD axions, the potential is determined entirely by the symmetry breaking scale $f_a$ since the mass and symmetry breaking scale are related by
\begin{equation}
    m \sim 6\times 10^{-10}\,\mathrm{eV}\left(\frac{10^{16}\, \mathrm{GeV}}{f_a}\right).
    \label{eq:mfRelation}
\end{equation} 
Thus the self-interaction $\lambda$ and the mass are not independent parameters.
For generic string theory axions, the potential is also a periodic function of the field with period $2\pi f_a$, so it has the same expansion to fourth order as in Eq.\ (\ref{eq:fourthorder}). However, the mass and axion decay constant are not related as in Eq.\ (\ref{eq:mfRelation}), so the potential has two independent parameters in this case.

The density of axions is extremely high, compared to the characteristic particle volume set by the de~Broglie wavelength $\lambda_\mathrm{dBr}$. For example in our galaxy \cite{SelfInteractionSign} the density of QCD axions is estimated to be 
\begin{equation}
    n_\mathrm{gal}\sim\frac{\rho_\mathrm{gal}}{m}\sim10^{14} \,\mathrm{cm}^{-3}.
\end{equation} The de~Broglie wavelength of virialized particles in an axion minicluster depends on the size of the minicluster, but for axions in our galaxy we have
\begin{equation}
    \lambda_\mathrm{dBr} \sim \frac{1}{mv} \sim 10^4 \,\mathrm{cm}.
\end{equation}
Thus the occupancy number is
\begin{equation}
    \mathcal N \sim n_\mathrm{gal} \lambda_\mathrm{dBr} ^3 \sim 10^{26}.
    \label{eq:occupancy}
\end{equation}

Under these high occupancy conditions, the coherent state axion dynamics can be approximated by a classical non-relativistic field. 
The approximation as a classical field is valid because the quantum fluctuations in a coherent state depend inversely on the occupancy number. If we expand
\begin{equation}
    \phi = \langle \hat \phi \rangle +\delta \hat \phi,
\end{equation}
where $\delta\hat\phi$ is the quantum fluctuations of $\phi$ about the mean field, then
\begin{equation}
    \delta\hat\phi \sim \frac{\phi}{\sqrt {\mathcal N}}.
\end{equation}
Equivalently, the timescale on which the quantum evolution of the field differs from the classical mean-field evolution is extremely long, orders of magnitude greater than the lifetime of the universe \cite{Dvali2018}.

Since the field is well approximated by a classical field, we can write down a classical action that couples the field to gravity. We consider only minimal coupling,
\begin{equation}
    S = \int \D^4 x\, \sqrt{-g}\left[ g^{\mu\nu}\partial_\mu\phi\partial_\nu\phi-\frac{1}{2}m^2\phi^2 - \lambda \phi^4\right],
\end{equation}
where $g^{\mu\nu}$ is the metric tensor. 
The classical equations of motion for this action are the Euler-Lagrange equations,
\begin{equation}
    \frac{1}{\sqrt{-g}}\partial_\mu \left(\sqrt{-g}g^{\mu\nu}
    \partial_\nu \phi\right) -m^2\phi -\lambda \phi^3 = 0.
\end{equation}
In the Newtonian limit, the metric takes the form
\begin{equation}
    \D s^2 = -\left(1+2U\right)\D t^2 + a(t)\left(1-2U \right)\D \mathbf r ^2,
\end{equation}
where $U$ is the Newtonian gravitational potential (to linear order) and $a$ is the Hubble scale factor.
This results in an equation of motion for $\phi$
\begin{equation}
    \ddot\phi - \frac{1-4U}{a(t)^2} \Delta \phi +3H\dot\phi -4\dot U \dot\phi + (1+2U) (m^2 \phi+\lambda \phi^3)=0,
    \label{eq:NewtonianGR}
\end{equation}
where $\dot\phi$ is $\partial\phi/\partial t$ and the Hubble parameter is $H=\dot a/a$.

Finally, we take the non-relativistic limit by writing the real scalar field $\phi$ in terms of a complex scalar field $\psi$ as
\begin{equation}
    \phi=\frac{1}{\sqrt{2m}}\left[\e^{-imt}\psi + \e^{imt}\psi^* \right].
    \label{eq:nonrel}
\end{equation}
In the non-relativistic limit, the phase factors $\e^{\pm imt}$ oscillate rapidly and when we substitute Eq.\ (\ref{eq:nonrel}) into Eq.\ (\ref{eq:NewtonianGR}), we can drop all terms with such a phase factor since they lead to only subdominant correction \cite{NamjooGuthKaiser,SalehianNamjooKaiser}. The result is the equations of motion for the complex field $\psi$, known as the Gross-Pitaevskii-Poisson equations (GPP) or the Nonlinear Schr\"odinger-Poisson equations
\begin{equation}\begin{aligned}
    i\partial_t \psi &= -\Delta \psi/2m + U\psi +\lambda |\psi|^2 \psi \\
    \Delta U &= 4\pi G m^2 (|\psi|^2 - n).
    \label{eq:Poisson}
\end{aligned}\end{equation}
In the above Poisson equation, we subtract the mean density $n$ for consistency \cite{JeansSwindleMath,JeansSwindleElectric}.
Here, $U$ is the Newtonian gravitational potential, $\lambda$ is the self-coupling, and $n$ is the average density of axions in an axion minicluster.

\subsection{Wigner distribution}

Recently,  Levkov, Panin and Tkachev \cite{Levkov2018} gave an argument a statistical ensemble of axions evolving under their self-gravity (without self-interactions) could not be treated as a standard Boltzmann collision process. They showed that since gravitational interactions are long range and interactions between distant axions are significant, the mean free path of the axions $a_\mathrm{gr}$ is very small with respect to $n^{1/3}$ since long-range fluctuations are important. I.e., we have a dimensionless ratio 
\begin{equation}
    a_\mathrm{gr} n^{1/3} \gg 1.
\end{equation}
This implies that a Boltzmann collision process is inappropriate for modeling the gravitational interactions: the particles are too dense to treat collisions as a process involving only two particles.

We provide an additional argument that the Boltzmann collision integral is not valid, even for the short range self-interactions. Even though the mean free path of self-interactions is small with respect to $n^{1/3}$, the de~Broglie wavelength is not, and we have another dimensionless ratio of length scales,
 \begin{equation}
     \lambda_\mathrm{dBr} n^{1/3} \gg 1,
 \end{equation}
This is a restatement of Eq.\ (\ref{eq:occupancy}) in terms of a dimensionless ratio of length scales. It tells us the quantum occupancy number $\mathcal N$ is high. 

This high occupancy number implies that the evolution of the statistical ensemble cannot be described by a standard Boltzmann equation because the axions cannot be localized to a definite position and momentum in phase space. Rather than describing an ensemble of particles by a phase space density, we can describe the ensemble by the Wigner function,
\begin{equation}
    f(\mathbf{x,p},t) = \int \D \mathbf y \,
    \e^{-i \mathbf{p\cdot y}}\langle \psi^*(\mathbf{x}+\frac{\mathbf y}{2})\psi(\mathbf{x}+\frac{\mathbf y}{2}) \rangle.
    \label{eq:Wigner}
\end{equation}
The Wigner function is the closest mathematical object we have to a phase space description for an ensemble of waves. In the appropriate limit, when the occupancy number becomes low, it recovers the properties of a positive-valued probability density function of particles. In our case where the occupancy number is high, the Wigner function reflects the inability to localize particles by taking negative values on regions of phase space whose size is on the order of $\hbar$ (i.e., on length scales set by the de~Broglie wavelength and momentum scales set by the characteristic momentum). The negative values obtained by the Wigner function are the result of interference of the waves, a phenomenon that is neglected in a classical particle description. As a result, the Wigner function has been used to study quantum properties scalar fields during inflation \cite{MartinVennin2016,MartinVennin2017}. 

The standard Boltzmann collision integral, which is developed for a localized collision of two or more particles, is not suited to describe the evolution of the Wigner function for the reasons stated above. Instead, we can systematically develop a kinetic equation by evolving the Wigner function by the GPP equations in Eq.\ (\ref{eq:Poisson}),
\begin{equation}
    \frac{\D f}{\D t} = 
    2\,\mathrm{Im}\int \D \mathbf y \,
    \e^{-i \mathbf{p\cdot y}}\left < \psi^*_+ \psi_-\U(\mathbf{x}+\frac{\mathbf y}{2})\right >,
    \label{eq:dfdt}
\end{equation}
which relates the evolution of the Wigner function to a four-point correlation function. ($U_\mathrm{tot}$ is defined in Eq.\ \ref{eq:Utot}. For a derivation of this equation of motion see Appendix A.) Here, we have simplified notation by denoting \begin{equation}
    \psi_\pm = \psi (\mathbf x \pm \frac{\mathbf y}{2}).
    \label{eq:subscripts}
\end{equation}
The total derivative is
\begin{equation}
    \frac{\D f}{\D t}=\frac{\partial f}{\partial t} + \frac{\mathbf p}{m} \cdot \nabla_\mathbf{x} f.
\end{equation}

By estimating the size of the four-point function, we obtain physical quantities such as the timescale for relaxation into the BEC state. Many numerical methods have been developed to study the evolution of the Wigner distribution in phase space \cite{Eberhardt2020}.

\section{Evolution of Wigner function}
\subsection{Equation of motion}
The gravitational potential $U(\mathbf x)$ is a functional of second order in $\psi$, the same order as the potential for nonlinear interactions, $\lambda|\psi|^2$. These combine to form a single potential $U_\mathrm{tot}$ and we rewrite the Gross-Pitaevskii-Poisson system as
\begin{equation}
    i\partial_t \psi = -\Delta \psi/2m + U_\mathrm{tot}\psi,
    \label{eq:EOM}
\end{equation}
where
\begin{equation}\begin{aligned}
    \U(\mathbf x) &= U(\mathbf x) + \lambda |\psi(\mathbf x)|^2 \\
    &= \int \D \mathbf x' \left[4 \pi G m^2\Delta^{-1} _\mathbf{x-x'} + \lambda \delta(\mathbf{x'-x}) \right]
    |\psi(\mathbf x')|^2 \\
    &\qquad-\int \D \mathbf x' 4 \pi G m^2 n \Delta^{-1} _\mathbf{x-x'},
    \label{eq:Utot}
\end{aligned}\end{equation}
and $\Delta ^{-1} _\mathbf{x-x'}$ is the Green's function for the Poisson equation,
\begin{equation}
    \Delta^{-1} _\mathbf{x-x'} = \frac{1}{4\pi |\mathbf{x-x'}|}.
\end{equation}

This can be expanded through Wick's theorem,
\begin{equation}\begin{aligned}
    \left< \psi_1 \psi^* _2 \psi_3 \psi^* _4 \right> = 
    \left<\psi_1 \psi^* _2 \right>
    \left<\psi_3 \psi^* _4 \right>
    +\left<\psi_1 \psi^* _4 \right>
    \left<\psi_3 \psi^* _2 \right> \\
    +\left< \psi_1 \psi^* _2 \psi_3 \psi^* _4 \right>_\mathrm{conn},
    \label{eq:Wick}
\end{aligned}\end{equation}
Here the subscripts refer to the spatial arguments of the fields in the four-point function, the first two terms are Wick contractions, and the last term is the connected correlation function, which is nonzero whenever the distribution is not Gaussian. 

We assume that the initial distribution of the field in an axion minicluster is Gaussian, with randomly distributed phases. This is appropriate for an uncorrelated, but gravitationally bound system like axion miniclusters immediately after their formation through the Kibble mechanism \cite{Fairbairn2018}. 
When the initial state is a Gaussian distribution, the connected correlation function in Eq.\ (\ref{eq:Wick}) vanishes. As the ensemble evolves, the interactions cause non-Gaussianities to develop and the connected correlations grow at rates set by $\lambda$ and $G$, as these are the coefficients of the nonlinear terms in the GPP equations.

Expanding the four-point function in Eq.\ (\ref{eq:dfdt}) we obtain three factors from the Wick expansion in Eq.\ (\ref{eq:Wick}),
\begin{equation}
    \frac{\D f}{\D t} = F_1 + F_2 + I(f),
\end{equation}
where $F_1$, $F_2$ are the contributions from the Wick contractions,
\begin{equation}\begin{aligned}
    F_1 &= 2 \, \mathrm{Im} \int_\mathbf{y} \, \e^{-i\mathbf p \cdot \mathbf y} 
    \left<\psi_+\psi^* _-\right> \left<\U(\mathbf x-\frac{\mathbf y}{2})\right> \\
   &=  2 \, \mathrm{Im} \,\left< U_\mathrm{tot} 
   (\mathbf x-\frac{i}{2}\nabla_\mathbf{p} )\right> f(\mathbf{x,p}),
\end{aligned}\end{equation}
\begin{equation}\begin{aligned}
    F_2 
    &= \frac{2}{(2\pi)^6} \mathrm{Im} \int_\mathbf{qq'yy'}\, 
    \left[ 4\pi Gm^2\Delta ^{-1} _\mathbf{y} +\lambda \delta(\mathbf y) \right] \\ 
    &\times
    \e^{i(\mathbf{y'\cdot q-y\cdot q'})} f(\mathbf x +\frac{\mathbf y}{2},\mathbf{p+q})f(\mathbf x + \frac{\mathbf y '}
    {2},\mathbf{p+q}'),
\end{aligned}\end{equation}
To save space we have indicated the variables of integration as subscripts, so e.g., $\int_\mathbf{y}=\int \D \mathbf{y}$. The scattering integral $I(f)$ depends on the connected correlator and will be addressed in Sec. IIIB.

In the case of a 
Gaussian ensemble of random waves, all three of these factors vanish. The scattering integral $I(f)$ is proportional to the connected correlations, so it vanishes because the initial distribution is Gaussian. In the terms $F_1$ and $F_2$, the Wigner functions, the Poisson Green's function, and the delta function are even in $\mathbf {y,y'}$, but the imaginary part of the integral selects the odd component of the potential. Thus the integration over the spatial coordinates $\mathbf {y,y'}$ causes these terms to vanish as well. Under these initial conditions, the distribution of axions is initially static.

In order to obtain any timescale for the evolution of the ensemble of waves, we need to look at the second derivative of the Wigner function. We will differentiate the four-point function as before using Eq.\ (\ref{eq:EOM}), and expand the resulting six-point function through Wick's theorem. However, because the gravitational and self interactions operate on different distance scales, we must evaluate this second derivative for the two interaction terms in a different way. We can do this because the correlation function is linear, and the terms proportional to $\lambda$ and $G$ can be separated. 





\subsection{Landau scattering integral}
The connected correlation function in the scattering integral is linear, so the terms governed by $\lambda$ and $G$ can be separated.
\begin{equation}\begin{aligned}
    I(f) &= I_\lambda + I_G , \\
    I_\lambda &= 2 \,\mathrm{Im} \int_\mathbf{y} \,
    e^{-i\mathbf p \cdot \mathbf y} \lambda
    \left< \psi_+ \psi^* _- 
    \psi_+ \psi^* _+\right> _\mathrm{conn} \\
   I_G &= 2 \,\mathrm{Im} \int_\mathbf{y} \,
    e^{-i\mathbf p \cdot \mathbf y}
    \left< \psi_+ \psi^* _- U\right> _\mathrm{conn}.
    \label{eq:scattering}
\end{aligned}\end{equation}
The term dependent on the gravitational potential $U$ contains long range interactions. In the context of plasmas with Coulomb interactions, Landau first noted that it is dominated by fluctuations at long distances (compared to the de~Broglie wavelength in this case) \cite{Liftschitz}. More recently in Ref. \cite{Levkov2018}, Landau's analysis was adapted to Newtonian gravity. 

Near $\mathbf x$ the potential has a multipole expansion
\begin{equation}\begin{aligned}
    U(\mathbf x +\mathbf y/2) &= 
    U(\mathbf x)+\frac{\mathbf y}{2}\cdot \nabla U(\mathbf x) + O(y^2) \\
    &= U(\mathbf x) + \frac{\mathbf y}{2}\cdot \int_{\mathbf x'} \nabla_\mathbf{x} \Delta^{-1} _{\mathbf{x-x}'} |\psi(\mathbf x')|^2 + O(y^2),
    \label{eq:multipole}
\end{aligned}\end{equation}
where $y=\left|\mathbf y\right|$. In an axion minicluster of radius $R$, the the field and the potential are nearly homogeneous on scales much shorter than $R$. So when $y\ll R$ we can truncate the multipole expansion at first order. Now in the integrals that follow, $U$ appears next to correlation functions of the form $\langle\psi(\mathbf x+\mathbf y/2)\psi^*(\mathbf x-\mathbf y/2)\rangle$, so the integrand is largest when the distance between the fields, $y$, is not much bigger than the correlation length $1/mv$. But in the kinetic regime
\begin{equation}
1/mv\ll R,
\end{equation}
so the multipole expansion of $U$ is valid to lowest order. Also, the equation of motion depends only on the part of $U$ that is odd in $\mathbf y$, so we get
\begin{equation}
     I_G(f) = i \int_\mathbf{x'y} \e^{-i\mathbf p\cdot \mathbf y} \langle \psi_+ \psi^* _- \psi_{\mathbf x'} \psi_{\mathbf x'}\rangle\, \mathbf y \cdot \nabla_\mathbf{x} \Delta^{-1} _{\mathbf{x-x}'}.
     \label{eq:ScatteringIntFirstOrder}
\end{equation}

Following Landau, we write the gravitational scattering integral as a diffusion process in phase space, in terms of a Landau Flux,
\begin{equation}
    \mathbf s = \frac{1}{(2\pi)^3} \int_\mathbf{x'p'} \mathcal F^{\mathbf{x'p'}} _{\mathbf{xp}} \nabla_\mathbf{x} \left[4\pi Gm^2\Delta ^{-1} _\mathbf{x-x'}\right],
    \label{eq:LandauFlux}
\end{equation}
where $\mathcal F$ is determined by the four-point connected correlator
\begin{equation}
    \mathcal F^{\mathbf{x'p'}} _\mathbf{xp} = 
    \int_\mathbf{yy'}
    e^{-i(\mathbf{py+p'y'})} \left<
    \psi_+ \psi^* _-
    {\psi _+}' {\psi^*  _- }'
    \right>_\mathrm{conn}.
\end{equation}
In this equation, $\psi_\pm$ are defined by Eq.\ (\ref{eq:subscripts}), while we also introduce the shorthand,
\begin{equation}
    \psi' _\pm = \psi(\mathbf x' \pm \frac{\mathbf y '}{2})
\end{equation}
(throughout this paper, primes are shorthand for the arguments of functions, not derivatives). With these definitions, we see
\begin{equation}
    I_G(f)=-\nabla_\mathbf{p} \cdot\mathbf s.
\end{equation}

An evolution equation for the four-point function $\mathcal F$ can be obtained from the equations of motion Eq.\ (\ref{eq:EOM}) as before,
\begin{equation}\begin{aligned}
    \frac{\D\mathcal F}{\D t} &=
    2\,\mathrm{Im}\int_\mathbf{yy'} \e^{-i\mathbf y\cdot\mathbf p -i \mathbf y'\cdot \mathbf p'} \\
    &\times\left< \psi_+ \psi^* _- {\psi_+}' {\psi^* _-}'
    \left(U_\mathrm{tot}(\mathbf x+\frac{\mathbf y}{2})+U_\mathrm{tot}(\mathbf x'+\frac{\mathbf y'}{2}) \right)\right>.
\end{aligned}\end{equation}
This is a six-point function which we can expand into Wick contractions. It also has a connected component which we will neglect since it introduces additional factors of $\lambda$ and $G$. We again use the multipole expansion to lowest order since the integral is dominated by short separations $y,y'\ll R$. Finally, we solve the ODE to obtain an expression for the Landau flux valid at time $t$,
\begin{align}
    s_i &= \int_\mathbf{p'} \Pi_{ij}(\mathbf u) \left(
    f'^2 \partial_{p_j} f - f^2 \partial_{p'_j} f'\right), \\
    \mathbf u &= (\mathbf{p'-p})/m, \\
    \Pi_{ij}(\mathbf u) &= \int \D t' \D\mathbf y \,
    \partial_i \left[4\pi Gm^2\Delta ^{-1} _\mathbf{y} \right]
    \\ &\qquad\times
    \partial_j \left[4\pi Gm^2\Delta ^{-1} _{\mathbf{y+u}t'} +\lambda\delta(\mathbf{y+u}t')  \right].
    \label{eq:flux}
\end{align}
In these equations, $f'=f(\mathbf x',\mathbf p')$ while $f=f(\mathbf x,\mathbf p)$, and we use Einstein summation over repeated indices.

Due to the logarithmic divergence from the Poisson Green's function, the integral over $t'$ must be regulated at both long and short time scales. The axion minicluster has a radius $R$ beyond which the axion field vanishes. Thus the long-time cutoff is $R/v$. Because the diffusion process is sensitive only to fluctuations at long distance, there is also a short-time cutoff $1/(mv^2)$, since the axion field lacks fluctuations at scales less than the de~Broglie wavelength. This completely suppresses the contribution from the self-interactions, which operate only at short distance scales.

Physically, the relaxation rate due to gravity grows with time as fluctuations at further distances begin to interact. However, it does not grow without bound, since the minicluster has a finite size $R$. For times $t \gtrsim R/v$ we can apply these short- and long-time cutoffs to the integral in Eq.\ (\ref{eq:flux}) to obtain the relaxation rate in this regime. As a result the integral depends on the Coulomb logarithm
\begin{equation}
    \Lambda=\log(mvR).
\end{equation}
The condensation timescale due to gravity is the inverse of this rate.

From this expression for the Landau flux, we can estimate the rate of change for the scattering integral due to gravitational interactions. I.e. we have 
\begin{equation}\begin{aligned}
    \frac{\D f}{\D t} &\sim I_\lambda + I_G \\
    I_G &\sim f/\tau_G,
    \label{eq:IG}
\end{aligned}\end{equation}
where
\begin{equation}
    \tau_G = \frac{\sqrt{2} m v^6}{12 \pi^3 G^2 n^2 \Lambda}.
\end{equation}

\subsection{Relaxation from self-coupling}
The self-interactions are not long range, so the scattering cannot be treated as a diffusion process in phase space like the scattering caused by gravitational interactions. We differentiate the equation of motion for the Wigner function in Eq.\ (\ref{eq:dfdt}), replacing $U_\mathrm{tot}$ with $\lambda |\psi|^2$ in that equation, since we have already treated the gravitational interactions in the previous section. This results in a six-point function which we consider to leading order in $\lambda$ and $G$, (therefore neglecting its connected correlations). The Wick contractions yield six contributions. In the case of a homogeneous ensemble of random waves, most of these factors vanish. But we are left with one non-vanishing contribution,
\begin{equation}\begin{aligned}
    \frac{\D^2 f}{\D t^2} &= 8\lambda^2 \int_\mathbf{y} \e^{-i\mathbf {p\cdot y}} \langle \psi_+\psi^* _- \rangle
    \langle \psi_- \psi^* _- \rangle^2 \\
    &\sim 8\lambda^2 \langle n\rangle^2 f(\mathbf{x,p}).
\end{aligned}\end{equation}
We recover a relaxation rate associated with the self interactions, 
\begin{equation}
    \frac{\D I_\lambda}{\D t} \sim \gamma f,
    \label{eq:dIdt}
\end{equation}
\begin{equation}
    \gamma = 8\lambda^2 n^2.
\end{equation}

Unlike the gravitational relaxation rate, the rate due to self-interactions does not depend on the distance between fluctuations, since the self-interactions are local. The relaxation rate does not grow with time until the connected correlations become significant. Instead we have a relation where $\D^2 f/\D t^2$ is directly proportional $f$, with proportionality constant $\gamma$. The constant $\gamma$ has units of $\mathrm s ^{-2}$, so to obtain the relaxation timescale from self-interactions, we take a square-root,
\begin{equation}
    \tau_\lambda =\frac{1}{\sqrt{\gamma}}=\frac{1}{\sqrt{8}|\lambda| n}.
\end{equation}

\section{Discussion}
There are two qualitative differences between the two timescales for Bose condensation that arise from the non-local nature of the gravitational interactions. First, the gravitational relaxation time depends explicitly on the size $R$ of the axion minicluster and on the axion's de~Broglie wavelength $1/(mv)$, through the Coulomb logarithm $\Lambda=\log(mvR)$, while the self-interaction relaxation time does not. This is a straightforward consequence of non-locality. The coupling ``constant" of gravitational interactions
\begin{equation}
    Gm^2/k
\end{equation}
depends on the momentum $k$, so these two natural distance scales appear as cutoffs in the logarithmic integral in Eq.\ (\ref{eq:flux}).

Second, the gravitational relaxation rate is proportional to $G^2$, while the self-interaction relaxation rate is only proportional to $\lambda$. This is also a consequence of non-locality. For the local self-interactions, we obtained a relation between the Wigner function and its second derivative in Eq.\ (\ref{eq:dIdt}). The rate $\gamma$ in that equation is proportional to $\lambda^2$, but because this is an expression for the second derivative, we must take a square root to obtain the characteristic timescale associated with this process, resulting in a timescale that is inversely proportional to $\lambda$. 

For the non-local gravitational interactions, it works out differently. Because the integral over $t'$ in Eq.\ (\ref{eq:flux}) does not go all the way to $t$ but is regulated by the short- and long-time cutoffs, we do not get a relation between the Wigner function and its second derivative. Instead we have a relation between the Wigner function and its first derivative, and we do not need to take the square root of the rate in Eq.\ (\ref{eq:IG}), so the timescale associated with this process is inversely proportional to $G^2$.

These qualitative differences between the two timescales, as well as the difference in the strength of the coupling constants, leads to significantly longer relaxation times due to self-interactions than due to gravity. We find that for QCD axions, the relaxation timescale for gravity is substantially shorter than the timescale for self-coupling, but not to such an extreme degree as has been previously reported. Since $\lambda$ only appears to first order in the relaxation rate while $G$ appears to second order, the small self-coupling strength does not increase the relaxation time as strongly. We find
\begin{equation}\begin{aligned}
    \tau_G &\sim \frac{v^6}{G^2 \rho_\mathrm{gal} \Lambda}\left(6\times 10^{-10}\,\mathrm{eV}\right)^3\left(\frac{10^{16}\,\mathrm{GeV}}{f_a}\right)^3\\
    &\sim 10^{17}\,\mathrm{s},
\end{aligned}\end{equation}
\begin{equation}\begin{aligned}
    \tau_\lambda &\sim \frac{1}{\rho_\mathrm{gal} f_a ^2} \left(6\times 10^{-10}\,\mathrm{eV}\right)^3\left(\frac{10^{16}\,\mathrm{GeV}}{f_a}\right)^3 \\
    &\sim 10^{22}\,\mathrm s.
\end{aligned}\end{equation}
In this case, Bose stars can form just within the lifetime of the universe due to their self-gravitation, while the self-interactions are too weak to have any effect during the formation process.

For ultralight scalar dark matter, the de~Broglie wavelength can be comparable to the size of the minicluster, $mvR\sim 1$ which strongly affects the gravitational relaxation time sensitive to the Coulomb logarithm $\Lambda$. Moreover the mass and self-interaction are not determined by a simple relation like Eq.\ (\ref{eq:mfRelation}) for QCD axions, so there are more parameters which can vary. Assuming that the Coulomb logarithm is $O(1)$, and taking the masses and interactions suggested by cosmological constraints in  Ref. \cite{CosmologicalConstraints}, we find that the condensation can occur much faster, though self-gravitation still dominates,
\begin{equation}
    \tau_G \sim 100 \,\mathrm{s},
\end{equation}
\begin{equation}
    \tau_\lambda \sim 10^{17}\,\mathrm{s}.
\end{equation}

When both gravity and self-interactions are present, the relaxation rate is simply the sum of the two rates, since at lowest order in $\lambda$ and $G$ there are no cross terms. Thus the total relaxation time is
\begin{equation}
    \tau_\mathrm{tot} \sim \frac{2\tau_\lambda\tau_G}{\tau_\lambda+\sqrt{\tau_\lambda ^2 + 4\tau_G ^2}}.
\end{equation}
When either timescale vastly exceeds the other, this reduces to the more familiar form,
\begin{equation}
    \tau_\mathrm{tot} \sim \frac{\tau_\lambda \tau_G}{\tau_\lambda+\tau_G}.
\end{equation}
As we have seen, the gravitational relaxation typically occurs much faster, so this expression reduces further to
\begin{equation}
    \tau_\mathrm{total}\sim\tau_G.
\end{equation}
This proves that the formation process of Bose stars is dominated by gravitational interactions. By the time self-interaction have an effect on the fields evolution, gravity has already caused the field to condense.

Finally, we note that while these calculations show that self-interactions of strength predicted for the QCD axion or most other scalar dark matter play a negligible role during the formation of the Bose-Einstein condensate, they can still play an important role in the phenomenology of the Bose stars. For example, 
Ref. \cite{SelfInteractionSign} showed the sign of the self-interactions can determine whether long-range correlations are possible, with such correlations impossible under attractive self-interactions. 
Ref. \cite{Chavanis2011} showed that the scattering length of self-coupling determines the mass-radius relation for Bose stars, as well as the maximum mass for which a stable equilibrium state exists. When this critical mass is exceeded, the axion star collapses and a number of phenomena can occur when the axions scatter under self-interactions \cite{ChavanisCollapse,EbyCollapse,LevkovCollapse}.
Finally, we note that recent studies have shown that there is a second branch of solutions to the GPP equations known as ``dense axion stars" in which self-interactions are significant and the full potential of Eq.\ (\ref{axionpotential}) is needed \cite{BraatenDense,Chavanis2018}. Whether this state is the result of the collapse process of overcritical dilute axion stars is currently unknown.

\subsection*{Acknowledgments}
We wish to thank Matthew Buckley, Priya Natarajan, Ed Copeland, Mustafa Amin, Richard Easther, Djuna Croon, Yanzhi Zhang, Andrew Eberhardt, Arka Banerjee, David Kaiser, and Noah Glennon for helpful discussions. CPW would like to thank all workers who made this research possible, especially those at the University of New Hampshire, the Aspen Center for Physics, which is supported by National Science Foundation (NSF) grant PHY-1607611, and the Kavli Institute for Theoretical Physics, where this research was supported in part by the NSF under Grant No. NSF PHY-1748958. CPW's contributions to this project were supported by DOE Grant DE-SC0020220. KK was partially supported by NSF CAREER Award DMS-1254791 and a Simons Sabbatical Fellowship. This paper honors the memory of Aiyana Stanley-Jones.

\subsection*{Appendix A: Equation of motion for Wigner function}
In this appendix, we justify Eq.\ (\ref{eq:dfdt}) by deriving the equation of motion for the Wigner function.

The equation of motion for the Wigner function $f$ is
\begin{equation}\begin{aligned}
    \frac{\D f}{\D t} &= -\int_\mathbf{y} \e^{-i\mathbf p\cdot\mathbf y} i \left(\langle\psi_+\psi_- ^* U(\mathbf x +\mathbf y/2)\rangle\right.\\
   &\qquad\qquad -\left.
    \langle\psi_+\psi_- ^* U(\mathbf x -\mathbf y/2)\rangle \right) \\
    &=-2i\int_\mathbf{y} \e^{-i\mathbf p\cdot\mathbf y} \langle\psi_+\psi_- ^* U^\mathrm{odd} (\mathbf x +\mathbf y/2)\rangle
\end{aligned}\end{equation}
where $\psi_\pm$ are as defined in Eq.\ (\ref{eq:subscripts}) and we define $U^\mathrm{odd}$ ($U^\mathrm{even})$ as the part of $U(\mathbf x+\mathbf y/2)$ which is odd (even) in $\mathbf y$.
Now let us define a new quantity 
\begin{equation}\begin{aligned}
    \mathcal A &= \int_\mathbf{y} \e^{-i\mathbf p\cdot\mathbf y} \langle\psi_+\psi_- ^* U (\mathbf x +\mathbf y/2)\rangle \\
    &= \int_\mathbf{y} \e^{-i\mathbf p\cdot\mathbf y}
    \left(\langle\psi_+\psi_- ^* U^\mathrm{even} (\mathbf x +\mathbf y/2)\rangle\right.\\
    &\qquad\qquad+\left.\langle\psi_+\psi_- ^* U^\mathrm{odd} (\mathbf x +\mathbf y/2)\rangle\right)
\end{aligned}\end{equation}
and show that $2\,\mathrm{Im}\mathcal A$ agrees with $\D f/\D t$ above. We do this in three steps. First, conjugate:
\begin{equation}\begin{aligned}
    \mathcal A^* &= \int_\mathbf{y} \e^{i\mathbf p\cdot\mathbf y} \left(\langle\psi_-\psi_+ ^* U^\mathrm{even} (\mathbf x +\mathbf y/2)\rangle\right. \\
    &\qquad\qquad+\left.\langle\psi_-\psi_+ ^* U^\mathrm{odd} (\mathbf x +\mathbf y/2)\rangle\right).
\end{aligned}\end{equation}
Next, change variables $\mathbf y\rightarrow -\mathbf y$:
\begin{equation}\begin{aligned}
    \mathcal A^* &= \int_\mathbf{y} \e^{-i\mathbf p\cdot\mathbf y} \left(\langle\psi_+\psi_- ^* U^\mathrm{even} (\mathbf x -\mathbf y/2)\rangle\right. \\
    &\qquad\qquad+\left.\langle\psi_+\psi_- ^* U^\mathrm{odd} (\mathbf x -\mathbf y/2)\rangle\right).
\end{aligned}\end{equation}
Finally, rewrite $U(\mathbf x-\mathbf y/2)$ in terms of $U(\mathbf x+\mathbf y/2)$ and combine with $\mathcal A$:
\begin{equation}\begin{aligned}
    2\,\mathrm{Im}\mathcal A &= \frac{\mathcal A-\mathcal A^*}{i} \\
    &= -2i \int_\mathbf{y} \e^{-i\mathbf p\cdot\mathbf y} \langle\psi_+\psi_- ^* U^\mathrm{odd} (\mathbf x +\mathbf y/2)\rangle.
\end{aligned}\end{equation}
This shows that $\D f/\D t$ is given by Eq.\ (\ref{eq:dfdt}).

\bibliographystyle{apsrev4-1}
\bibliography{TonyAxion.bib}

\begin{thebibliography}{66}%
\makeatletter
\providecommand \@ifxundefined [1]{%
 \@ifx{#1\undefined}
}%
\providecommand \@ifnum [1]{%
 \ifnum #1\expandafter \@firstoftwo
 \else \expandafter \@secondoftwo
 \fi
}%
\providecommand \@ifx [1]{%
 \ifx #1\expandafter \@firstoftwo
 \else \expandafter \@secondoftwo
 \fi
}%
\providecommand \natexlab [1]{#1}%
\providecommand \enquote  [1]{``#1''}%
\providecommand \bibnamefont  [1]{#1}%
\providecommand \bibfnamefont [1]{#1}%
\providecommand \citenamefont [1]{#1}%
\providecommand \href@noop [0]{\@secondoftwo}%
\providecommand \href [0]{\begingroup \@sanitize@url \@href}%
\providecommand \@href[1]{\@@startlink{#1}\@@href}%
\providecommand \@@href[1]{\endgroup#1\@@endlink}%
\providecommand \@sanitize@url [0]{\catcode `\\12\catcode `\$12\catcode
  `\&12\catcode `\#12\catcode `\^12\catcode `\_12\catcode `\%12\relax}%
\providecommand \@@startlink[1]{}%
\providecommand \@@endlink[0]{}%
\providecommand \url  [0]{\begingroup\@sanitize@url \@url }%
\providecommand \@url [1]{\endgroup\@href {#1}{\urlprefix }}%
\providecommand \urlprefix  [0]{URL }%
\providecommand \Eprint [0]{\href }%
\providecommand \doibase [0]{http://dx.doi.org/}%
\providecommand \selectlanguage [0]{\@gobble}%
\providecommand \bibinfo  [0]{\@secondoftwo}%
\providecommand \bibfield  [0]{\@secondoftwo}%
\providecommand \translation [1]{[#1]}%
\providecommand \BibitemOpen [0]{}%
\providecommand \bibitemStop [0]{}%
\providecommand \bibitemNoStop [0]{.\EOS\space}%
\providecommand \EOS [0]{\spacefactor3000\relax}%
\providecommand \BibitemShut  [1]{\csname bibitem#1\endcsname}%
\let\auto@bib@innerbib\@empty
\bibitem [{\citenamefont {{Planck Collaboration}}\ \emph
  {et~al.}(2018)\citenamefont {{Planck Collaboration}}, \citenamefont {Aghanim}
  \emph {et~al.}}]{PLANCK2018}%
  \BibitemOpen
  \bibfield  {author} {\bibinfo {author} {\bibnamefont {{Planck
  Collaboration}}}, \bibinfo {author} {\bibfnamefont {N.}~\bibnamefont
  {Aghanim}},  \emph {et~al.},\ }\href@noop {} {\bibfield  {journal} {\bibinfo
  {journal} {arXiv Preprints}\ } (\bibinfo {year} {2018})},\ \Eprint
  {http://arxiv.org/abs/arXiv:1807.06209} {arXiv:1807.06209 [astro-ph.CO]}
  \BibitemShut {NoStop}%
\bibitem [{\citenamefont {Weinberg}\ \emph {et~al.}(2015)\citenamefont
  {Weinberg}, \citenamefont {Bullock}, \citenamefont {Governato}, \citenamefont
  {Kuzio~de Naray},\ and\ \citenamefont {Peter}}]{WeinbergControversies2015}%
  \BibitemOpen
  \bibfield  {author} {\bibinfo {author} {\bibfnamefont {D.~H.}\ \bibnamefont
  {Weinberg}}, \bibinfo {author} {\bibfnamefont {J.~S.}\ \bibnamefont
  {Bullock}}, \bibinfo {author} {\bibfnamefont {F.}~\bibnamefont {Governato}},
  \bibinfo {author} {\bibfnamefont {R.}~\bibnamefont {Kuzio~de Naray}}, \ and\
  \bibinfo {author} {\bibfnamefont {A.~H.}\ \bibnamefont {Peter}},\ }\href
  {\doibase 10.1073/pnas.1308716112} {\bibfield  {journal} {\bibinfo  {journal}
  {Proceedings of the National Academy of Sciences}\ }\textbf {\bibinfo
  {volume} {112}},\ \bibinfo {pages} {12249} (\bibinfo {year}
  {2015})}\BibitemShut {NoStop}%
\bibitem [{\citenamefont {Moore}(1994)}]{CDMproblems}%
  \BibitemOpen
  \bibfield  {author} {\bibinfo {author} {\bibfnamefont {B.}~\bibnamefont
  {Moore}},\ }\href@noop {} {\bibfield  {journal} {\bibinfo  {journal}
  {Nature}\ }\textbf {\bibinfo {volume} {370}},\ \bibinfo {pages} {629}
  (\bibinfo {year} {1994})}\BibitemShut {NoStop}%
\bibitem [{\citenamefont {{Papastergis, E.}}\ \emph {et~al.}(2015)\citenamefont
  {{Papastergis, E.}}, \citenamefont {{Giovanelli, R.}}, \citenamefont
  {{Haynes, M. P.}},\ and\ \citenamefont {{Shankar, F.}}}]{TooBigToFail}%
  \BibitemOpen
  \bibfield  {author} {\bibinfo {author} {\bibnamefont {{Papastergis, E.}}},
  \bibinfo {author} {\bibnamefont {{Giovanelli, R.}}}, \bibinfo {author}
  {\bibnamefont {{Haynes, M. P.}}}, \ and\ \bibinfo {author} {\bibnamefont
  {{Shankar, F.}}},\ }\href {\doibase 10.1051/0004-6361/201424909} {\bibfield
  {journal} {\bibinfo  {journal} {A\&A}\ }\textbf {\bibinfo {volume} {574}},\
  \bibinfo {pages} {A113} (\bibinfo {year} {2015})}\BibitemShut {NoStop}%
\bibitem [{\citenamefont {Avila-Reese}\ \emph {et~al.}(2001)\citenamefont
  {Avila-Reese}, \citenamefont {Colin}, \citenamefont {Valenzuela},
  \citenamefont {D'Onghia},\ and\ \citenamefont {Firmani}}]{AvilaReese2001}%
  \BibitemOpen
  \bibfield  {author} {\bibinfo {author} {\bibfnamefont {V.}~\bibnamefont
  {Avila-Reese}}, \bibinfo {author} {\bibfnamefont {P.}~\bibnamefont {Colin}},
  \bibinfo {author} {\bibfnamefont {O.}~\bibnamefont {Valenzuela}}, \bibinfo
  {author} {\bibfnamefont {E.}~\bibnamefont {D'Onghia}}, \ and\ \bibinfo
  {author} {\bibfnamefont {C.}~\bibnamefont {Firmani}},\ }\href {\doibase
  10.1086/322411} {\bibfield  {journal} {\bibinfo  {journal} {The Astrophysical
  Journal}\ }\textbf {\bibinfo {volume} {559}},\ \bibinfo {pages} {516}
  (\bibinfo {year} {2001})}\BibitemShut {NoStop}%
\bibitem [{\citenamefont {Kamionkowski}\ and\ \citenamefont
  {Liddle}(2000)}]{Kamionkowski2000}%
  \BibitemOpen
  \bibfield  {author} {\bibinfo {author} {\bibfnamefont {M.}~\bibnamefont
  {Kamionkowski}}\ and\ \bibinfo {author} {\bibfnamefont {A.~R.}\ \bibnamefont
  {Liddle}},\ }\href {\doibase 10.1103/PhysRevLett.84.4525} {\bibfield
  {journal} {\bibinfo  {journal} {Phys. Rev. Lett.}\ }\textbf {\bibinfo
  {volume} {84}},\ \bibinfo {pages} {4525} (\bibinfo {year}
  {2000})}\BibitemShut {NoStop}%
\bibitem [{\citenamefont {Spergel}\ and\ \citenamefont
  {Steinhardt}(2000)}]{Spergel2000}%
  \BibitemOpen
  \bibfield  {author} {\bibinfo {author} {\bibfnamefont {D.~N.}\ \bibnamefont
  {Spergel}}\ and\ \bibinfo {author} {\bibfnamefont {P.~J.}\ \bibnamefont
  {Steinhardt}},\ }\href {\doibase 10.1103/PhysRevLett.84.3760} {\bibfield
  {journal} {\bibinfo  {journal} {Phys. Rev. Lett.}\ }\textbf {\bibinfo
  {volume} {84}},\ \bibinfo {pages} {3760} (\bibinfo {year}
  {2000})}\BibitemShut {NoStop}%
\bibitem [{\citenamefont {Governato}\ \emph {et~al.}(2010)\citenamefont
  {Governato}, \citenamefont {Brook}, \citenamefont {Mayer}, \citenamefont
  {Brooks}, \citenamefont {Rhee}, \citenamefont {Wadsley}, \citenamefont
  {Jonsson}, \citenamefont {Willman}, \citenamefont {Stinson}, \citenamefont
  {Quinn},\ and\ \citenamefont {Madau}}]{Governato2010}%
  \BibitemOpen
  \bibfield  {author} {\bibinfo {author} {\bibfnamefont {F.}~\bibnamefont
  {Governato}}, \bibinfo {author} {\bibfnamefont {C.}~\bibnamefont {Brook}},
  \bibinfo {author} {\bibfnamefont {L.}~\bibnamefont {Mayer}}, \bibinfo
  {author} {\bibfnamefont {A.}~\bibnamefont {Brooks}}, \bibinfo {author}
  {\bibfnamefont {G.}~\bibnamefont {Rhee}}, \bibinfo {author} {\bibfnamefont
  {J.}~\bibnamefont {Wadsley}}, \bibinfo {author} {\bibfnamefont
  {P.}~\bibnamefont {Jonsson}}, \bibinfo {author} {\bibfnamefont
  {B.}~\bibnamefont {Willman}}, \bibinfo {author} {\bibfnamefont
  {G.}~\bibnamefont {Stinson}}, \bibinfo {author} {\bibfnamefont
  {T.}~\bibnamefont {Quinn}}, \ and\ \bibinfo {author} {\bibfnamefont
  {P.}~\bibnamefont {Madau}},\ }\href@noop {} {\bibfield  {journal} {\bibinfo
  {journal} {Nature}\ }\textbf {\bibinfo {volume} {463}},\ \bibinfo {pages}
  {203} (\bibinfo {year} {2010})}\BibitemShut {NoStop}%
\bibitem [{\citenamefont {Buckley}\ and\ \citenamefont
  {Peter}(2018)}]{BuckleyPeter2018}%
  \BibitemOpen
  \bibfield  {author} {\bibinfo {author} {\bibfnamefont {M.~R.}\ \bibnamefont
  {Buckley}}\ and\ \bibinfo {author} {\bibfnamefont {A.~H.~G.}\ \bibnamefont
  {Peter}},\ }\href@noop {} {\bibfield  {journal} {\bibinfo  {journal} {Physics
  Reports}\ }\textbf {\bibinfo {volume} {761}},\ \bibinfo {pages} {1} (\bibinfo
  {year} {2018})}\BibitemShut {NoStop}%
\bibitem [{\citenamefont {Peccei}\ and\ \citenamefont
  {Quinn}(1977)}]{PecceiQuinn}%
  \BibitemOpen
  \bibfield  {author} {\bibinfo {author} {\bibfnamefont {R.~D.}\ \bibnamefont
  {Peccei}}\ and\ \bibinfo {author} {\bibfnamefont {H.~R.}\ \bibnamefont
  {Quinn}},\ }\href {\doibase 10.1103/PhysRevLett.38.1440} {\bibfield
  {journal} {\bibinfo  {journal} {Phys. Rev. Lett.}\ }\textbf {\bibinfo
  {volume} {38}},\ \bibinfo {pages} {1440} (\bibinfo {year}
  {1977})}\BibitemShut {NoStop}%
\bibitem [{\citenamefont {Weinberg}(1978)}]{Weinberg1978}%
  \BibitemOpen
  \bibfield  {author} {\bibinfo {author} {\bibfnamefont {S.}~\bibnamefont
  {Weinberg}},\ }\href {\doibase 10.1103/PhysRevLett.40.223} {\bibfield
  {journal} {\bibinfo  {journal} {Phys. Rev. Lett.}\ }\textbf {\bibinfo
  {volume} {40}},\ \bibinfo {pages} {223} (\bibinfo {year} {1978})}\BibitemShut
  {NoStop}%
\bibitem [{\citenamefont {Wilczek}(1978)}]{Wilczek1978}%
  \BibitemOpen
  \bibfield  {author} {\bibinfo {author} {\bibfnamefont {F.}~\bibnamefont
  {Wilczek}},\ }\href {\doibase 10.1103/PhysRevLett.40.279} {\bibfield
  {journal} {\bibinfo  {journal} {Phys. Rev. Lett.}\ }\textbf {\bibinfo
  {volume} {40}},\ \bibinfo {pages} {279} (\bibinfo {year} {1978})}\BibitemShut
  {NoStop}%
\bibitem [{\citenamefont {Dine}\ \emph {et~al.}(1981)\citenamefont {Dine},
  \citenamefont {Fischler},\ and\ \citenamefont
  {Srednicki}}]{DineFischlerSrednicki}%
  \BibitemOpen
  \bibfield  {author} {\bibinfo {author} {\bibfnamefont {M.}~\bibnamefont
  {Dine}}, \bibinfo {author} {\bibfnamefont {W.}~\bibnamefont {Fischler}}, \
  and\ \bibinfo {author} {\bibfnamefont {M.}~\bibnamefont {Srednicki}},\
  }\href@noop {} {\bibfield  {journal} {\bibinfo  {journal} {Physics Letters
  B}\ }\textbf {\bibinfo {volume} {104}},\ \bibinfo {pages} {199} (\bibinfo
  {year} {1981})}\BibitemShut {NoStop}%
\bibitem [{\citenamefont {Turner}(1986)}]{Turner1986}%
  \BibitemOpen
  \bibfield  {author} {\bibinfo {author} {\bibfnamefont {M.~S.}\ \bibnamefont
  {Turner}},\ }\href {\doibase 10.1103/PhysRevD.33.889} {\bibfield  {journal}
  {\bibinfo  {journal} {Phys. Rev. D}\ }\textbf {\bibinfo {volume} {33}},\
  \bibinfo {pages} {889} (\bibinfo {year} {1986})}\BibitemShut {NoStop}%
\bibitem [{\citenamefont {Sikivie}(2008)}]{SikivieNotes}%
  \BibitemOpen
  \bibfield  {author} {\bibinfo {author} {\bibfnamefont {P.}~\bibnamefont
  {Sikivie}},\ }\enquote {\bibinfo {title} {Axions. lecture notes in physics,
  vol 741.}}\ \ (\bibinfo  {publisher} {Springer, Berlin, Heidelberg},\
  \bibinfo {year} {2008})\ Chap.\ \bibinfo {chapter} {Axion
  Cosmology}\BibitemShut {NoStop}%
\bibitem [{\citenamefont {Arvanitaki}\ \emph {et~al.}(2010)\citenamefont
  {Arvanitaki}, \citenamefont {Dimopoulos}, \citenamefont {Dubovsky},
  \citenamefont {Kaloper},\ and\ \citenamefont
  {March-Russell}}]{StringAxiverse}%
  \BibitemOpen
  \bibfield  {author} {\bibinfo {author} {\bibfnamefont {A.}~\bibnamefont
  {Arvanitaki}}, \bibinfo {author} {\bibfnamefont {S.}~\bibnamefont
  {Dimopoulos}}, \bibinfo {author} {\bibfnamefont {S.}~\bibnamefont
  {Dubovsky}}, \bibinfo {author} {\bibfnamefont {N.}~\bibnamefont {Kaloper}}, \
  and\ \bibinfo {author} {\bibfnamefont {J.}~\bibnamefont {March-Russell}},\
  }\href {\doibase 10.1103/PhysRevD.81.123530} {\bibfield  {journal} {\bibinfo
  {journal} {Phys. Rev. D}\ }\textbf {\bibinfo {volume} {81}},\ \bibinfo
  {pages} {123530} (\bibinfo {year} {2010})}\BibitemShut {NoStop}%
\bibitem [{\citenamefont {Hui}\ \emph {et~al.}(2017)\citenamefont {Hui},
  \citenamefont {Ostriker}, \citenamefont {Tremaine},\ and\ \citenamefont
  {Witten}}]{WittenUltralight}%
  \BibitemOpen
  \bibfield  {author} {\bibinfo {author} {\bibfnamefont {L.}~\bibnamefont
  {Hui}}, \bibinfo {author} {\bibfnamefont {J.~P.}\ \bibnamefont {Ostriker}},
  \bibinfo {author} {\bibfnamefont {S.}~\bibnamefont {Tremaine}}, \ and\
  \bibinfo {author} {\bibfnamefont {E.}~\bibnamefont {Witten}},\ }\href
  {\doibase 10.1103/PhysRevD.95.043541} {\bibfield  {journal} {\bibinfo
  {journal} {Phys. Rev. D}\ }\textbf {\bibinfo {volume} {95}},\ \bibinfo
  {pages} {043541} (\bibinfo {year} {2017})}\BibitemShut {NoStop}%
\bibitem [{\citenamefont {Ringwald}(2012)}]{Ringwald2012}%
  \BibitemOpen
  \bibfield  {author} {\bibinfo {author} {\bibfnamefont {A.}~\bibnamefont
  {Ringwald}},\ }\href@noop {} {\bibfield  {journal} {\bibinfo  {journal}
  {Physics of the Dark Universe}\ }\textbf {\bibinfo {volume} {1}},\ \bibinfo
  {pages} {116} (\bibinfo {year} {2012})}\BibitemShut {NoStop}%
\bibitem [{\citenamefont {John~Preskill}(1983)}]{Preskill1983}%
  \BibitemOpen
  \bibfield  {author} {\bibinfo {author} {\bibfnamefont {F.~W.}\ \bibnamefont
  {John~Preskill}, \bibfnamefont {Mark B.~Wise}},\ }\href@noop {} {\bibfield
  {journal} {\bibinfo  {journal} {Physics Letters B}\ }\textbf {\bibinfo
  {volume} {120}},\ \bibinfo {pages} {127} (\bibinfo {year}
  {1983})}\BibitemShut {NoStop}%
\bibitem [{\citenamefont {Kim}\ and\ \citenamefont {Carosi}(2010)}]{Kim2010}%
  \BibitemOpen
  \bibfield  {author} {\bibinfo {author} {\bibfnamefont {J.~E.}\ \bibnamefont
  {Kim}}\ and\ \bibinfo {author} {\bibfnamefont {G.}~\bibnamefont {Carosi}},\
  }\href {\doibase 10.1103/RevModPhys.82.557} {\bibfield  {journal} {\bibinfo
  {journal} {Rev. Mod. Phys.}\ }\textbf {\bibinfo {volume} {82}},\ \bibinfo
  {pages} {557} (\bibinfo {year} {2010})}\BibitemShut {NoStop}%
\bibitem [{\citenamefont {Braine}\ and\ \citenamefont {others {[ADMX
  Collaboration]}}(2020)}]{ADMX2020}%
  \BibitemOpen
  \bibfield  {author} {\bibinfo {author} {\bibfnamefont {T.}~\bibnamefont
  {Braine}}\ and\ \bibinfo {author} {\bibnamefont {others {[ADMX
  Collaboration]}}} (\bibinfo {collaboration} {ADMX Collaboration}),\ }\href
  {\doibase 10.1103/PhysRevLett.124.101303} {\bibfield  {journal} {\bibinfo
  {journal} {Phys. Rev. Lett.}\ }\textbf {\bibinfo {volume} {124}},\ \bibinfo
  {pages} {101303} (\bibinfo {year} {2020})}\BibitemShut {NoStop}%
\bibitem [{\citenamefont {Ouellet}\ \emph {et~al.}(2019)\citenamefont
  {Ouellet}, \citenamefont {Salemi}, \citenamefont {Foster}, \citenamefont
  {Henning}, \citenamefont {Bogorad}, \citenamefont {Conrad}, \citenamefont
  {Formaggio}, \citenamefont {Kahn}, \citenamefont {Minervini}, \citenamefont
  {Radovinsky}, \citenamefont {Rodd}, \citenamefont {Safdi}, \citenamefont
  {Thaler}, \citenamefont {Winklehner},\ and\ \citenamefont
  {Winslow}}]{ABRACADABRA2019}%
  \BibitemOpen
  \bibfield  {author} {\bibinfo {author} {\bibfnamefont {J.~L.}\ \bibnamefont
  {Ouellet}}, \bibinfo {author} {\bibfnamefont {C.~P.}\ \bibnamefont {Salemi}},
  \bibinfo {author} {\bibfnamefont {J.~W.}\ \bibnamefont {Foster}}, \bibinfo
  {author} {\bibfnamefont {R.}~\bibnamefont {Henning}}, \bibinfo {author}
  {\bibfnamefont {Z.}~\bibnamefont {Bogorad}}, \bibinfo {author} {\bibfnamefont
  {J.~M.}\ \bibnamefont {Conrad}}, \bibinfo {author} {\bibfnamefont {J.~A.}\
  \bibnamefont {Formaggio}}, \bibinfo {author} {\bibfnamefont {Y.}~\bibnamefont
  {Kahn}}, \bibinfo {author} {\bibfnamefont {J.}~\bibnamefont {Minervini}},
  \bibinfo {author} {\bibfnamefont {A.}~\bibnamefont {Radovinsky}}, \bibinfo
  {author} {\bibfnamefont {N.~L.}\ \bibnamefont {Rodd}}, \bibinfo {author}
  {\bibfnamefont {B.~R.}\ \bibnamefont {Safdi}}, \bibinfo {author}
  {\bibfnamefont {J.}~\bibnamefont {Thaler}}, \bibinfo {author} {\bibfnamefont
  {D.}~\bibnamefont {Winklehner}}, \ and\ \bibinfo {author} {\bibfnamefont
  {L.}~\bibnamefont {Winslow}},\ }\href {\doibase
  10.1103/PhysRevLett.122.121802} {\bibfield  {journal} {\bibinfo  {journal}
  {Phys. Rev. Lett.}\ }\textbf {\bibinfo {volume} {122}},\ \bibinfo {pages}
  {121802} (\bibinfo {year} {2019})}\BibitemShut {NoStop}%
\bibitem [{\citenamefont {Abel}\ \emph {et~al.}(2020)\citenamefont {Abel} \emph
  {et~al.}}]{NeutronDipole2020}%
  \BibitemOpen
  \bibfield  {author} {\bibinfo {author} {\bibfnamefont {C.}~\bibnamefont
  {Abel}} \emph {et~al.},\ }\href {\doibase 10.1103/PhysRevLett.124.081803}
  {\bibfield  {journal} {\bibinfo  {journal} {Phys. Rev. Lett.}\ }\textbf
  {\bibinfo {volume} {124}},\ \bibinfo {pages} {081803} (\bibinfo {year}
  {2020})}\BibitemShut {NoStop}%
\bibitem [{\citenamefont {Lee}\ and\ \citenamefont {Pang}(1992)}]{Lee1992}%
  \BibitemOpen
  \bibfield  {author} {\bibinfo {author} {\bibfnamefont {T.~D.}\ \bibnamefont
  {Lee}}\ and\ \bibinfo {author} {\bibfnamefont {Y.}~\bibnamefont {Pang}},\
  }\href@noop {} {\bibfield  {journal} {\bibinfo  {journal} {Physics Reports}\
  }\textbf {\bibinfo {volume} {221}},\ \bibinfo {pages} {251} (\bibinfo {year}
  {1992})}\BibitemShut {NoStop}%
\bibitem [{\citenamefont {Jetzer}(1992)}]{Jetzer1992}%
  \BibitemOpen
  \bibfield  {author} {\bibinfo {author} {\bibfnamefont {P.}~\bibnamefont
  {Jetzer}},\ }\href@noop {} {\bibfield  {journal} {\bibinfo  {journal}
  {Physics Reports}\ }\textbf {\bibinfo {volume} {220}},\ \bibinfo {pages}
  {163} (\bibinfo {year} {1992})}\BibitemShut {NoStop}%
\bibitem [{\citenamefont {Kolb}\ and\ \citenamefont
  {Tkachev}(1993)}]{Kolb1993}%
  \BibitemOpen
  \bibfield  {author} {\bibinfo {author} {\bibfnamefont {E.~W.}\ \bibnamefont
  {Kolb}}\ and\ \bibinfo {author} {\bibfnamefont {I.~I.}\ \bibnamefont
  {Tkachev}},\ }\href {\doibase 10.1103/PhysRevLett.71.3051} {\bibfield
  {journal} {\bibinfo  {journal} {Phys. Rev. Lett.}\ }\textbf {\bibinfo
  {volume} {71}},\ \bibinfo {pages} {3051} (\bibinfo {year}
  {1993})}\BibitemShut {NoStop}%
\bibitem [{\citenamefont {Guth}\ \emph {et~al.}(2015)\citenamefont {Guth},
  \citenamefont {Hertzberg},\ and\ \citenamefont
  {Prescod-Weinstein}}]{SelfInteractionSign}%
  \BibitemOpen
  \bibfield  {author} {\bibinfo {author} {\bibfnamefont {A.~H.}\ \bibnamefont
  {Guth}}, \bibinfo {author} {\bibfnamefont {M.~P.}\ \bibnamefont {Hertzberg}},
  \ and\ \bibinfo {author} {\bibfnamefont {C.}~\bibnamefont
  {Prescod-Weinstein}},\ }\href {\doibase 10.1103/PhysRevD.92.103513}
  {\bibfield  {journal} {\bibinfo  {journal} {Phys. Rev. D}\ }\textbf {\bibinfo
  {volume} {92}},\ \bibinfo {pages} {103513} (\bibinfo {year}
  {2015})}\BibitemShut {NoStop}%
\bibitem [{\citenamefont {Semikoz}\ and\ \citenamefont
  {Tkachev}(1997)}]{SemikozTkachev}%
  \BibitemOpen
  \bibfield  {author} {\bibinfo {author} {\bibfnamefont {D.~V.}\ \bibnamefont
  {Semikoz}}\ and\ \bibinfo {author} {\bibfnamefont {I.~I.}\ \bibnamefont
  {Tkachev}},\ }\href {\doibase 10.1103/PhysRevD.55.489} {\bibfield  {journal}
  {\bibinfo  {journal} {Phys. Rev. D}\ }\textbf {\bibinfo {volume} {55}},\
  \bibinfo {pages} {489} (\bibinfo {year} {1997})}\BibitemShut {NoStop}%
\bibitem [{\citenamefont {Khlebnikov}(2000)}]{Khlebnikov2000}%
  \BibitemOpen
  \bibfield  {author} {\bibinfo {author} {\bibfnamefont {S.}~\bibnamefont
  {Khlebnikov}},\ }\href {\doibase 10.1103/PhysRevD.62.043519} {\bibfield
  {journal} {\bibinfo  {journal} {Phys. Rev. D}\ }\textbf {\bibinfo {volume}
  {62}},\ \bibinfo {pages} {043519} (\bibinfo {year} {2000})}\BibitemShut
  {NoStop}%
\bibitem [{\citenamefont {Sikivie}\ and\ \citenamefont
  {Yang}(2009)}]{Sikivie2009}%
  \BibitemOpen
  \bibfield  {author} {\bibinfo {author} {\bibfnamefont {P.}~\bibnamefont
  {Sikivie}}\ and\ \bibinfo {author} {\bibfnamefont {Q.}~\bibnamefont {Yang}},\
  }\href {\doibase 10.1103/PhysRevLett.103.111301} {\bibfield  {journal}
  {\bibinfo  {journal} {Phys. Rev. Lett.}\ }\textbf {\bibinfo {volume} {103}},\
  \bibinfo {pages} {111301} (\bibinfo {year} {2009})}\BibitemShut {NoStop}%
\bibitem [{\citenamefont {Erken}\ \emph {et~al.}(2012)\citenamefont {Erken},
  \citenamefont {Sikivie}, \citenamefont {Tam},\ and\ \citenamefont
  {Yang}}]{Erken2012}%
  \BibitemOpen
  \bibfield  {author} {\bibinfo {author} {\bibfnamefont {O.}~\bibnamefont
  {Erken}}, \bibinfo {author} {\bibfnamefont {P.}~\bibnamefont {Sikivie}},
  \bibinfo {author} {\bibfnamefont {H.}~\bibnamefont {Tam}}, \ and\ \bibinfo
  {author} {\bibfnamefont {Q.}~\bibnamefont {Yang}},\ }\href {\doibase
  10.1103/PhysRevD.85.063520} {\bibfield  {journal} {\bibinfo  {journal} {Phys.
  Rev. D}\ }\textbf {\bibinfo {volume} {85}},\ \bibinfo {pages} {063520}
  (\bibinfo {year} {2012})}\BibitemShut {NoStop}%
\bibitem [{\citenamefont {Ruffini}\ and\ \citenamefont
  {Bonazzola}(1969)}]{RuffiniBonazzola}%
  \BibitemOpen
  \bibfield  {author} {\bibinfo {author} {\bibfnamefont {R.}~\bibnamefont
  {Ruffini}}\ and\ \bibinfo {author} {\bibfnamefont {S.}~\bibnamefont
  {Bonazzola}},\ }\href {\doibase 10.1103/PhysRev.187.1767} {\bibfield
  {journal} {\bibinfo  {journal} {Phys. Rev.}\ }\textbf {\bibinfo {volume}
  {187}},\ \bibinfo {pages} {1767} (\bibinfo {year} {1969})}\BibitemShut
  {NoStop}%
\bibitem [{\citenamefont {Chavanis}(2011)}]{Chavanis2011}%
  \BibitemOpen
  \bibfield  {author} {\bibinfo {author} {\bibfnamefont {P.-H.}\ \bibnamefont
  {Chavanis}},\ }\href {\doibase 10.1103/PhysRevD.84.043531} {\bibfield
  {journal} {\bibinfo  {journal} {Phys. Rev. D}\ }\textbf {\bibinfo {volume}
  {84}},\ \bibinfo {pages} {043531} (\bibinfo {year} {2011})}\BibitemShut
  {NoStop}%
\bibitem [{\citenamefont {Barranco}\ and\ \citenamefont
  {Bernal}(2011)}]{Barranco2011}%
  \BibitemOpen
  \bibfield  {author} {\bibinfo {author} {\bibfnamefont {J.}~\bibnamefont
  {Barranco}}\ and\ \bibinfo {author} {\bibfnamefont {A.}~\bibnamefont
  {Bernal}},\ }\href {\doibase 10.1103/PhysRevD.83.043525} {\bibfield
  {journal} {\bibinfo  {journal} {Phys. Rev. D}\ }\textbf {\bibinfo {volume}
  {83}},\ \bibinfo {pages} {043525} (\bibinfo {year} {2011})}\BibitemShut
  {NoStop}%
\bibitem [{\citenamefont {Eby}\ \emph {et~al.}(2019)\citenamefont {Eby},
  \citenamefont {Leembruggen}, \citenamefont {Street}, \citenamefont
  {Suranyi},\ and\ \citenamefont {Wijewardhana}}]{EbyGlobalView}%
  \BibitemOpen
  \bibfield  {author} {\bibinfo {author} {\bibfnamefont {J.}~\bibnamefont
  {Eby}}, \bibinfo {author} {\bibfnamefont {M.}~\bibnamefont {Leembruggen}},
  \bibinfo {author} {\bibfnamefont {L.}~\bibnamefont {Street}}, \bibinfo
  {author} {\bibfnamefont {P.}~\bibnamefont {Suranyi}}, \ and\ \bibinfo
  {author} {\bibfnamefont {L.}~\bibnamefont {Wijewardhana}},\ }\href {\doibase
  10.1103/PhysRevD.100.063002} {\bibfield  {journal} {\bibinfo  {journal}
  {Phys. Rev. D}\ }\textbf {\bibinfo {volume} {100}},\ \bibinfo {pages}
  {063002} (\bibinfo {year} {2019})}\BibitemShut {NoStop}%
\bibitem [{\citenamefont {Eby}\ \emph {et~al.}(2016{\natexlab{a}})\citenamefont
  {Eby}, \citenamefont {Kouvaris}, \citenamefont {Nielsen},\ and\ \citenamefont
  {Wijewardhana}}]{EbySelfInteraction}%
  \BibitemOpen
  \bibfield  {author} {\bibinfo {author} {\bibfnamefont {J.}~\bibnamefont
  {Eby}}, \bibinfo {author} {\bibfnamefont {C.}~\bibnamefont {Kouvaris}},
  \bibinfo {author} {\bibfnamefont {N.~G.}\ \bibnamefont {Nielsen}}, \ and\
  \bibinfo {author} {\bibfnamefont {L.}~\bibnamefont {Wijewardhana}},\
  }\href@noop {} {\bibfield  {journal} {\bibinfo  {journal} {Journal of High
  Energy Physics}\ }\textbf {\bibinfo {volume} {2016}},\ \bibinfo {pages} {28}
  (\bibinfo {year} {2016}{\natexlab{a}})}\BibitemShut {NoStop}%
\bibitem [{\citenamefont {Eby}\ \emph {et~al.}(2015)\citenamefont {Eby},
  \citenamefont {Suranyi}, \citenamefont {Vaz},\ and\ \citenamefont
  {Wijewardhana}}]{EbyInfrared}%
  \BibitemOpen
  \bibfield  {author} {\bibinfo {author} {\bibfnamefont {J.}~\bibnamefont
  {Eby}}, \bibinfo {author} {\bibfnamefont {P.}~\bibnamefont {Suranyi}},
  \bibinfo {author} {\bibfnamefont {C.}~\bibnamefont {Vaz}}, \ and\ \bibinfo
  {author} {\bibfnamefont {L.}~\bibnamefont {Wijewardhana}},\ }\href@noop {}
  {\bibfield  {journal} {\bibinfo  {journal} {Journal of High Energy Physics}\
  }\textbf {\bibinfo {volume} {2015}},\ \bibinfo {pages} {80} (\bibinfo {year}
  {2015})}\BibitemShut {NoStop}%
\bibitem [{\citenamefont {Eby}\ \emph {et~al.}(2016{\natexlab{b}})\citenamefont
  {Eby}, \citenamefont {Suranyi},\ and\ \citenamefont
  {Wijewardhana}}]{EbyLifetime}%
  \BibitemOpen
  \bibfield  {author} {\bibinfo {author} {\bibfnamefont {J.}~\bibnamefont
  {Eby}}, \bibinfo {author} {\bibfnamefont {P.}~\bibnamefont {Suranyi}}, \ and\
  \bibinfo {author} {\bibfnamefont {L.}~\bibnamefont {Wijewardhana}},\
  }\href@noop {} {\bibfield  {journal} {\bibinfo  {journal} {Modern Physics
  Letters A}\ }\textbf {\bibinfo {volume} {31}},\ \bibinfo {pages} {1650090}
  (\bibinfo {year} {2016}{\natexlab{b}})}\BibitemShut {NoStop}%
\bibitem [{\citenamefont {Eby}\ \emph {et~al.}(2018)\citenamefont {Eby},
  \citenamefont {Leembruggen}, \citenamefont {Street}, \citenamefont
  {Suranyi},\ and\ \citenamefont {Wijewardhana}}]{EbyApprox}%
  \BibitemOpen
  \bibfield  {author} {\bibinfo {author} {\bibfnamefont {J.}~\bibnamefont
  {Eby}}, \bibinfo {author} {\bibfnamefont {M.}~\bibnamefont {Leembruggen}},
  \bibinfo {author} {\bibfnamefont {L.}~\bibnamefont {Street}}, \bibinfo
  {author} {\bibfnamefont {P.}~\bibnamefont {Suranyi}}, \ and\ \bibinfo
  {author} {\bibfnamefont {L.}~\bibnamefont {Wijewardhana}},\ }\href {\doibase
  10.1103/PhysRevD.98.123013} {\bibfield  {journal} {\bibinfo  {journal} {Phys.
  Rev. D}\ }\textbf {\bibinfo {volume} {98}},\ \bibinfo {pages} {123013}
  (\bibinfo {year} {2018})}\BibitemShut {NoStop}%
\bibitem [{\citenamefont {Braaten}\ \emph {et~al.}(2016)\citenamefont
  {Braaten}, \citenamefont {Mohapatra},\ and\ \citenamefont
  {Zhang}}]{BraatenDense}%
  \BibitemOpen
  \bibfield  {author} {\bibinfo {author} {\bibfnamefont {E.}~\bibnamefont
  {Braaten}}, \bibinfo {author} {\bibfnamefont {A.}~\bibnamefont {Mohapatra}},
  \ and\ \bibinfo {author} {\bibfnamefont {H.}~\bibnamefont {Zhang}},\ }\href
  {\doibase 10.1103/PhysRevLett.117.121801} {\bibfield  {journal} {\bibinfo
  {journal} {Phys. Rev. Lett.}\ }\textbf {\bibinfo {volume} {117}},\ \bibinfo
  {pages} {121801} (\bibinfo {year} {2016})}\BibitemShut {NoStop}%
\bibitem [{\citenamefont {Visinelli}\ \emph {et~al.}(2018)\citenamefont
  {Visinelli}, \citenamefont {Baum}, \citenamefont {Redondo}, \citenamefont
  {Freese},\ and\ \citenamefont {Wilczek}}]{Visinelli2018}%
  \BibitemOpen
  \bibfield  {author} {\bibinfo {author} {\bibfnamefont {L.}~\bibnamefont
  {Visinelli}}, \bibinfo {author} {\bibfnamefont {S.}~\bibnamefont {Baum}},
  \bibinfo {author} {\bibfnamefont {J.}~\bibnamefont {Redondo}}, \bibinfo
  {author} {\bibfnamefont {K.}~\bibnamefont {Freese}}, \ and\ \bibinfo {author}
  {\bibfnamefont {F.}~\bibnamefont {Wilczek}},\ }\href@noop {} {\bibfield
  {journal} {\bibinfo  {journal} {Physics Letters B}\ }\textbf {\bibinfo
  {volume} {777}},\ \bibinfo {pages} {64} (\bibinfo {year} {2018})}\BibitemShut
  {NoStop}%
\bibitem [{\citenamefont {Schive}\ \emph {et~al.}(2014)\citenamefont {Schive},
  \citenamefont {Chiueh},\ and\ \citenamefont {Broadhurst}}]{Schive2014}%
  \BibitemOpen
  \bibfield  {author} {\bibinfo {author} {\bibfnamefont {H.-Y.}\ \bibnamefont
  {Schive}}, \bibinfo {author} {\bibfnamefont {T.}~\bibnamefont {Chiueh}}, \
  and\ \bibinfo {author} {\bibfnamefont {T.}~\bibnamefont {Broadhurst}},\
  }\href@noop {} {\bibfield  {journal} {\bibinfo  {journal} {Nature Physics}\
  }\textbf {\bibinfo {volume} {10}},\ \bibinfo {pages} {496} (\bibinfo {year}
  {2014})}\BibitemShut {NoStop}%
\bibitem [{\citenamefont {Levkov}\ \emph {et~al.}(2018)\citenamefont {Levkov},
  \citenamefont {Panin},\ and\ \citenamefont {Tkachev}}]{Levkov2018}%
  \BibitemOpen
  \bibfield  {author} {\bibinfo {author} {\bibfnamefont {D.~G.}\ \bibnamefont
  {Levkov}}, \bibinfo {author} {\bibfnamefont {A.~G.}\ \bibnamefont {Panin}}, \
  and\ \bibinfo {author} {\bibfnamefont {I.~I.}\ \bibnamefont {Tkachev}},\
  }\href {\doibase 10.1103/PhysRevLett.121.151301} {\bibfield  {journal}
  {\bibinfo  {journal} {Phys. Rev. Lett.}\ }\textbf {\bibinfo {volume} {121}},\
  \bibinfo {pages} {151301} (\bibinfo {year} {2018})}\BibitemShut {NoStop}%
\bibitem [{\citenamefont {Marsh}(2016)}]{AxionCosmology}%
  \BibitemOpen
  \bibfield  {author} {\bibinfo {author} {\bibfnamefont {D.~J.~E.}\
  \bibnamefont {Marsh}},\ }\href@noop {} {\bibfield  {journal} {\bibinfo
  {journal} {Physics Reports}\ }\textbf {\bibinfo {volume} {643}},\ \bibinfo
  {pages} {1} (\bibinfo {year} {2016})}\BibitemShut {NoStop}%
\bibitem [{\citenamefont {Hogan}\ and\ \citenamefont {Rees}(1988)}]{HoganRees}%
  \BibitemOpen
  \bibfield  {author} {\bibinfo {author} {\bibfnamefont {C.}~\bibnamefont
  {Hogan}}\ and\ \bibinfo {author} {\bibfnamefont {M.}~\bibnamefont {Rees}},\
  }\href@noop {} {\bibfield  {journal} {\bibinfo  {journal} {Physics Letters
  B}\ }\textbf {\bibinfo {volume} {205}},\ \bibinfo {pages} {228} (\bibinfo
  {year} {1988})}\BibitemShut {NoStop}%
\bibitem [{\citenamefont {Nelson}\ and\ \citenamefont
  {Xiao}(2018)}]{Nelson2018}%
  \BibitemOpen
  \bibfield  {author} {\bibinfo {author} {\bibfnamefont {A.~E.}\ \bibnamefont
  {Nelson}}\ and\ \bibinfo {author} {\bibfnamefont {H.}~\bibnamefont {Xiao}},\
  }\href {\doibase 10.1103/PhysRevD.98.063516} {\bibfield  {journal} {\bibinfo
  {journal} {Phys. Rev. D}\ }\textbf {\bibinfo {volume} {98}},\ \bibinfo
  {pages} {063516} (\bibinfo {year} {2018})}\BibitemShut {NoStop}%
\bibitem [{\citenamefont {Kibble}(1976)}]{Kibble1976}%
  \BibitemOpen
  \bibfield  {author} {\bibinfo {author} {\bibfnamefont {T.}~\bibnamefont
  {Kibble}},\ }\href {\doibase 10.1088/0305-4470/9/8/029} {\bibfield  {journal}
  {\bibinfo  {journal} {Journal of Physics A: Mathematical and General}\
  }\textbf {\bibinfo {volume} {9}},\ \bibinfo {pages} {1387} (\bibinfo {year}
  {1976})}\BibitemShut {NoStop}%
\bibitem [{\citenamefont {Enander}\ \emph {et~al.}(2017)\citenamefont
  {Enander}, \citenamefont {Pargner},\ and\ \citenamefont
  {Schwetz}}]{MiniclusterSpectrum}%
  \BibitemOpen
  \bibfield  {author} {\bibinfo {author} {\bibfnamefont {J.}~\bibnamefont
  {Enander}}, \bibinfo {author} {\bibfnamefont {A.}~\bibnamefont {Pargner}}, \
  and\ \bibinfo {author} {\bibfnamefont {T.}~\bibnamefont {Schwetz}},\ }\href
  {\doibase 10.1088/1475-7516/2017/12/038} {\bibfield  {journal} {\bibinfo
  {journal} {Journal of Cosmology and Astroparticle Physics}\ }\textbf
  {\bibinfo {volume} {2017}},\ \bibinfo {pages} {038} (\bibinfo {year}
  {2017})}\BibitemShut {NoStop}%
\bibitem [{\citenamefont {Arvanitaki}\ \emph {et~al.}(2020)\citenamefont
  {Arvanitaki}, \citenamefont {Dimopoulos}, \citenamefont {Galanis},
  \citenamefont {Lehner}, \citenamefont {Thompson},\ and\ \citenamefont
  {Van~Tilburg}}]{Arvanitaki2020}%
  \BibitemOpen
  \bibfield  {author} {\bibinfo {author} {\bibfnamefont {A.}~\bibnamefont
  {Arvanitaki}}, \bibinfo {author} {\bibfnamefont {S.}~\bibnamefont
  {Dimopoulos}}, \bibinfo {author} {\bibfnamefont {M.}~\bibnamefont {Galanis}},
  \bibinfo {author} {\bibfnamefont {L.}~\bibnamefont {Lehner}}, \bibinfo
  {author} {\bibfnamefont {J.~O.}\ \bibnamefont {Thompson}}, \ and\ \bibinfo
  {author} {\bibfnamefont {K.}~\bibnamefont {Van~Tilburg}},\ }\href {\doibase
  10.1103/PhysRevD.101.083014} {\bibfield  {journal} {\bibinfo  {journal}
  {Phys. Rev. D}\ }\textbf {\bibinfo {volume} {101}},\ \bibinfo {pages}
  {083014} (\bibinfo {year} {2020})}\BibitemShut {NoStop}%
\bibitem [{\citenamefont {Hardy}(2017)}]{Hardy2017}%
  \BibitemOpen
  \bibfield  {author} {\bibinfo {author} {\bibfnamefont {E.}~\bibnamefont
  {Hardy}},\ }\href@noop {} {\bibfield  {journal} {\bibinfo  {journal} {Journal
  of High Energy Physics}\ }\textbf {\bibinfo {volume} {2017}},\ \bibinfo
  {pages} {46} (\bibinfo {year} {2017})}\BibitemShut {NoStop}%
\bibitem [{\citenamefont {Hayato~Fukunaga}(2016)}]{Fukunaga2020}%
  \BibitemOpen
  \bibfield  {author} {\bibinfo {author} {\bibfnamefont {Y.~U.}\ \bibnamefont
  {Hayato~Fukunaga}, \bibfnamefont {Naoya~Kitajima}},\ }\href@noop {}
  {\bibfield  {journal} {\bibinfo  {journal} {arXiv Preprints}\ } (\bibinfo
  {year} {2016})},\ \Eprint {http://arxiv.org/abs/2004.08929} {arXiv:2004.08929
  [astro-ph.CO]} \BibitemShut {NoStop}%
\bibitem [{\citenamefont {Dvali}\ and\ \citenamefont {Zell}(2018)}]{Dvali2018}%
  \BibitemOpen
  \bibfield  {author} {\bibinfo {author} {\bibfnamefont {G.}~\bibnamefont
  {Dvali}}\ and\ \bibinfo {author} {\bibfnamefont {S.}~\bibnamefont {Zell}},\
  }\href {\doibase 10.1088/1475-7516/2018/07/064} {\bibfield  {journal}
  {\bibinfo  {journal} {Journal of Cosmology and Astroparticle Physics}\
  }\textbf {\bibinfo {volume} {2018}},\ \bibinfo {pages} {064} (\bibinfo {year}
  {2018})}\BibitemShut {NoStop}%
\bibitem [{\citenamefont {Namjoo}\ \emph {et~al.}(2018)\citenamefont {Namjoo},
  \citenamefont {Guth},\ and\ \citenamefont {Kaiser}}]{NamjooGuthKaiser}%
  \BibitemOpen
  \bibfield  {author} {\bibinfo {author} {\bibfnamefont {M.~H.}\ \bibnamefont
  {Namjoo}}, \bibinfo {author} {\bibfnamefont {A.~H.}\ \bibnamefont {Guth}}, \
  and\ \bibinfo {author} {\bibfnamefont {D.~I.}\ \bibnamefont {Kaiser}},\
  }\href {\doibase 10.1103/PhysRevD.98.016011} {\bibfield  {journal} {\bibinfo
  {journal} {Phys. Rev. D}\ }\textbf {\bibinfo {volume} {98}},\ \bibinfo
  {pages} {016011} (\bibinfo {year} {2018})}\BibitemShut {NoStop}%
\bibitem [{\citenamefont {Salehian}\ \emph {et~al.}(2020)\citenamefont
  {Salehian}, \citenamefont {Namjoo},\ and\ \citenamefont
  {Kaiser}}]{SalehianNamjooKaiser}%
  \BibitemOpen
  \bibfield  {author} {\bibinfo {author} {\bibfnamefont {B.}~\bibnamefont
  {Salehian}}, \bibinfo {author} {\bibfnamefont {M.~H.}\ \bibnamefont
  {Namjoo}}, \ and\ \bibinfo {author} {\bibfnamefont {D.~I.}\ \bibnamefont
  {Kaiser}},\ }\href@noop {} {\bibfield  {journal} {\bibinfo  {journal}
  {Journal of High Energy Physics}\ }\textbf {\bibinfo {volume} {2020}},\
  \bibinfo {pages} {59} (\bibinfo {year} {2020})}\BibitemShut {NoStop}%
\bibitem [{\citenamefont {Kiessling}(2003)}]{JeansSwindleMath}%
  \BibitemOpen
  \bibfield  {author} {\bibinfo {author} {\bibfnamefont {M.~K.-H.}\
  \bibnamefont {Kiessling}},\ }\href@noop {} {\bibfield  {journal} {\bibinfo
  {journal} {Advances in Applied Mathematics}\ }\textbf {\bibinfo {volume}
  {31}},\ \bibinfo {pages} {132} (\bibinfo {year} {2003})}\BibitemShut
  {NoStop}%
\bibitem [{\citenamefont {Dabo}\ \emph {et~al.}(2008)\citenamefont {Dabo},
  \citenamefont {Kozinsky}, \citenamefont {Singh-Miller},\ and\ \citenamefont
  {Marzari}}]{JeansSwindleElectric}%
  \BibitemOpen
  \bibfield  {author} {\bibinfo {author} {\bibfnamefont {I.}~\bibnamefont
  {Dabo}}, \bibinfo {author} {\bibfnamefont {B.}~\bibnamefont {Kozinsky}},
  \bibinfo {author} {\bibfnamefont {N.~E.}\ \bibnamefont {Singh-Miller}}, \
  and\ \bibinfo {author} {\bibfnamefont {N.}~\bibnamefont {Marzari}},\ }\href
  {\doibase 10.1103/PhysRevB.77.115139} {\bibfield  {journal} {\bibinfo
  {journal} {Phys. Rev. B}\ }\textbf {\bibinfo {volume} {77}},\ \bibinfo
  {pages} {115139} (\bibinfo {year} {2008})}\BibitemShut {NoStop}%
\bibitem [{\citenamefont {Martin}\ and\ \citenamefont
  {Vennin}(2016)}]{MartinVennin2016}%
  \BibitemOpen
  \bibfield  {author} {\bibinfo {author} {\bibfnamefont {J.}~\bibnamefont
  {Martin}}\ and\ \bibinfo {author} {\bibfnamefont {V.}~\bibnamefont
  {Vennin}},\ }\href {\doibase 10.1103/PhysRevD.93.023505} {\bibfield
  {journal} {\bibinfo  {journal} {Phys. Rev. D}\ }\textbf {\bibinfo {volume}
  {93}},\ \bibinfo {pages} {023505} (\bibinfo {year} {2016})}\BibitemShut
  {NoStop}%
\bibitem [{\citenamefont {Martin}\ and\ \citenamefont
  {Vennin}(2017)}]{MartinVennin2017}%
  \BibitemOpen
  \bibfield  {author} {\bibinfo {author} {\bibfnamefont {J.}~\bibnamefont
  {Martin}}\ and\ \bibinfo {author} {\bibfnamefont {V.}~\bibnamefont
  {Vennin}},\ }\href {\doibase 10.1103/PhysRevD.96.063501} {\bibfield
  {journal} {\bibinfo  {journal} {Phys. Rev. D}\ }\textbf {\bibinfo {volume}
  {96}},\ \bibinfo {pages} {063501} (\bibinfo {year} {2017})}\BibitemShut
  {NoStop}%
\bibitem [{\citenamefont {Eberhardt}\ \emph {et~al.}(2020)\citenamefont
  {Eberhardt}, \citenamefont {Banerjee}, \citenamefont {Kopp},\ and\
  \citenamefont {Abel}}]{Eberhardt2020}%
  \BibitemOpen
  \bibfield  {author} {\bibinfo {author} {\bibfnamefont {A.}~\bibnamefont
  {Eberhardt}}, \bibinfo {author} {\bibfnamefont {A.}~\bibnamefont {Banerjee}},
  \bibinfo {author} {\bibfnamefont {M.}~\bibnamefont {Kopp}}, \ and\ \bibinfo
  {author} {\bibfnamefont {T.}~\bibnamefont {Abel}},\ }\href {\doibase
  10.1103/PhysRevD.101.043011} {\bibfield  {journal} {\bibinfo  {journal}
  {Phys. Rev. D}\ }\textbf {\bibinfo {volume} {101}},\ \bibinfo {pages}
  {043011} (\bibinfo {year} {2020})}\BibitemShut {NoStop}%
\bibitem [{\citenamefont {Fairbairn}\ \emph {et~al.}(2018)\citenamefont
  {Fairbairn}, \citenamefont {Marsh}, \citenamefont {Quevillon},\ and\
  \citenamefont {Rozier}}]{Fairbairn2018}%
  \BibitemOpen
  \bibfield  {author} {\bibinfo {author} {\bibfnamefont {M.}~\bibnamefont
  {Fairbairn}}, \bibinfo {author} {\bibfnamefont {D.~J.~E.}\ \bibnamefont
  {Marsh}}, \bibinfo {author} {\bibfnamefont {J.}~\bibnamefont {Quevillon}}, \
  and\ \bibinfo {author} {\bibfnamefont {S.}~\bibnamefont {Rozier}},\ }\href
  {\doibase 10.1103/PhysRevD.97.083502} {\bibfield  {journal} {\bibinfo
  {journal} {Phys. Rev. D}\ }\textbf {\bibinfo {volume} {97}},\ \bibinfo
  {pages} {083502} (\bibinfo {year} {2018})}\BibitemShut {NoStop}%
\bibitem [{\citenamefont {Lifshitz}\ and\ \citenamefont
  {Pitaevskij}(2005)}]{Liftschitz}%
  \BibitemOpen
  \bibfield  {author} {\bibinfo {author} {\bibfnamefont {E.~M.}\ \bibnamefont
  {Lifshitz}}\ and\ \bibinfo {author} {\bibfnamefont {L.~P.}\ \bibnamefont
  {Pitaevskij}},\ }\href@noop {} {\emph {\bibinfo {title} {Physical Kinetics,
  Landau and Lifschitz Course of Theoretical Physics Volume 10}}}\ (\bibinfo
  {publisher} {Elsevier},\ \bibinfo {year} {2005})\BibitemShut {NoStop}%
\bibitem [{\citenamefont {Li}\ \emph {et~al.}(2014)\citenamefont {Li},
  \citenamefont {Rindler-Daller},\ and\ \citenamefont
  {Shapiro}}]{CosmologicalConstraints}%
  \BibitemOpen
  \bibfield  {author} {\bibinfo {author} {\bibfnamefont {B.}~\bibnamefont
  {Li}}, \bibinfo {author} {\bibfnamefont {T.}~\bibnamefont {Rindler-Daller}},
  \ and\ \bibinfo {author} {\bibfnamefont {P.~R.}\ \bibnamefont {Shapiro}},\
  }\href {\doibase 10.1103/PhysRevD.89.083536} {\bibfield  {journal} {\bibinfo
  {journal} {Phys. Rev. D}\ }\textbf {\bibinfo {volume} {89}},\ \bibinfo
  {pages} {083536} (\bibinfo {year} {2014})}\BibitemShut {NoStop}%
\bibitem [{\citenamefont {Chavanis}(2016)}]{ChavanisCollapse}%
  \BibitemOpen
  \bibfield  {author} {\bibinfo {author} {\bibfnamefont {P.-H.}\ \bibnamefont
  {Chavanis}},\ }\href {\doibase 10.1103/PhysRevD.94.083007} {\bibfield
  {journal} {\bibinfo  {journal} {Phys. Rev. D}\ }\textbf {\bibinfo {volume}
  {94}},\ \bibinfo {pages} {083007} (\bibinfo {year} {2016})}\BibitemShut
  {NoStop}%
\bibitem [{\citenamefont {Eby}\ \emph {et~al.}(2016{\natexlab{c}})\citenamefont
  {Eby}, \citenamefont {Leembruggen}, \citenamefont {Suranyi},\ and\
  \citenamefont {Wijewardhana}}]{EbyCollapse}%
  \BibitemOpen
  \bibfield  {author} {\bibinfo {author} {\bibfnamefont {J.}~\bibnamefont
  {Eby}}, \bibinfo {author} {\bibfnamefont {M.}~\bibnamefont {Leembruggen}},
  \bibinfo {author} {\bibfnamefont {P.}~\bibnamefont {Suranyi}}, \ and\
  \bibinfo {author} {\bibfnamefont {L.}~\bibnamefont {Wijewardhana}},\
  }\href@noop {} {\bibfield  {journal} {\bibinfo  {journal} {Journal of High
  Energy Physics}\ }\textbf {\bibinfo {volume} {2016}},\ \bibinfo {pages} {66}
  (\bibinfo {year} {2016}{\natexlab{c}})}\BibitemShut {NoStop}%
\bibitem [{\citenamefont {Levkov}\ \emph {et~al.}(2017)\citenamefont {Levkov},
  \citenamefont {Panin},\ and\ \citenamefont {Tkachev}}]{LevkovCollapse}%
  \BibitemOpen
  \bibfield  {author} {\bibinfo {author} {\bibfnamefont {D.~G.}\ \bibnamefont
  {Levkov}}, \bibinfo {author} {\bibfnamefont {A.~G.}\ \bibnamefont {Panin}}, \
  and\ \bibinfo {author} {\bibfnamefont {I.~I.}\ \bibnamefont {Tkachev}},\
  }\href {\doibase 10.1103/PhysRevLett.118.011301} {\bibfield  {journal}
  {\bibinfo  {journal} {Phys. Rev. Lett.}\ }\textbf {\bibinfo {volume} {118}},\
  \bibinfo {pages} {011301} (\bibinfo {year} {2017})}\BibitemShut {NoStop}%
\bibitem [{\citenamefont {Chavanis}(2018)}]{Chavanis2018}%
  \BibitemOpen
  \bibfield  {author} {\bibinfo {author} {\bibfnamefont {P.-H.}\ \bibnamefont
  {Chavanis}},\ }\href {\doibase 10.1103/PhysRevD.98.023009} {\bibfield
  {journal} {\bibinfo  {journal} {Phys. Rev. D}\ }\textbf {\bibinfo {volume}
  {98}},\ \bibinfo {pages} {023009} (\bibinfo {year} {2018})}\BibitemShut
  {NoStop}%
\end{thebibliography}%

\end{document}